\documentclass[prd, twocolumn, nofootinbib, showpacs, superscriptaddress]{revtex4-1}
\usepackage[utf8]{inputenc}
\usepackage{graphicx}
\usepackage{color}
\usepackage{amsmath}
\usepackage{hyperref}
\usepackage{tabularx}
\usepackage{amsmath,amssymb}
\usepackage{float}

\usepackage[normalem]{ulem}
\newcommand{\be}{\begin{equation}}
\newcommand{\ee}{\end{equation}}
\newcommand{\bea}{\begin{eqnarray}}
\newcommand{\eea}{\end{eqnarray}}

\newcommand{\gapp}{\mathrel{\raise.3ex\hbox{$>$}\mkern-14mu \lower0.6ex\hbox{$\sim$}}}
\newcommand{\lapp}{\mathrel{\raise.3ex\hbox{$<$}\mkern-14mu \lower0.6ex\hbox{$\sim$}}}
\def\bbox{{\,\lower0.9pt\vbox{\hrule \hbox{\vrule height 0.2 cm
\hskip 0.2 cm \vrule  height 0.2 cm}\hrule}\,}}
\newcommand{\dd}{\mathrm d}

\usepackage{color}
\usepackage{ulem}
\usepackage[usenames,dvipsnames]{xcolor}

\begin{document}
\title{Quasi-normal modes of black holes in scalar-tensor theories with non-minimal derivative couplings}

\author{Ruifeng Dong}
\email[Email: ]{ruifengd@buffalo.edu}
\address{HEPCOS, Department of Physics, SUNY at Buffalo, Buffalo, NY 14260-1500}
\author{Jeremy Sakstein}
\email[Email: ]{sakstein@physics.upenn.edu}
\address{Department of Physics and Astronomy, Center for Particle Cosmology,
University of Pennsylvania, 209 S. 33rd St., Philadelphia, PA 19104, USA}
\author{Dejan Stojkovic}
\email[Email: ]{ds77@buffalo.edu}
\address{HEPCOS, Department of Physics, SUNY at Buffalo, Buffalo, NY 14260-1500}


\begin{abstract}
We study the quasi-normal modes of asymptotically anti-de Sitter black holes in a class of shift-symmetric Horndeski theories where a gravitational scalar is derivatively coupled to the Einstein tensor. The space-time differs from exact Schwarzschild-anti-de Sitter, resulting in a different effective potential for the quasi-normal modes and a different spectrum. We numerically compute this spectrum for a massless test scalar coupled both minimally to the metric, and non-minimally to the gravitational scalar. We find interesting differences from the Schwarzschild-anti-de Sitter black hole found in general relativity.

\end{abstract}


\maketitle

\section{Introduction}

The mysterious nature of dark energy \cite{Copeland:2006wr} has galvanized a recent theoretical study of alternative gravity theories as one potential driving mechanism for the acceleration of the cosmic expansion. The search for new and phenomenologically interesting theories has led to a proliferation of scalar-tensor extensions of general relativity (GR) \cite{Clifton:2011jh,Joyce:2014kja,Berti:2015itd,Koyama:2015vza,Burrage:2016bwy}. Many of these differ from the classical theories of modified gravity (such as Brans-Dickie) in that they include higher-derivative interactions, yet they are free of any Ostrogradski ghost instabilities because the equations of motion are second-order. These theories have received particular attention because they can self-accelerate cosmologically whilst simultaneously satisfying solar system tests of gravity by utilising the Vainshtein screening mechanism \cite{Vainshtein:1972sx,Nicolis:2008in,Kaloper:2011qc,Kimura:2011dc,Babichev:2013usa,Koyama:2013paa}, which uses non-linearities in the field equations to suppress deviations from GR. 

Any scalar-tensor theory that has second-order equations of motion falls into the class of theories first derived by Horndeski \cite{Horndeski:1974wa} and independently re-derived by \cite{Deffayet:2009wt,Deffayet:2009mn,Deffayet:2011gz}. This class is defined by four free functions of a the scalar $\varphi$ and its kinetic energy $X=-g^{\mu\nu}\partial_\mu\varphi\partial_\nu\varphi/2$ and a set of essential building blocks. Such an expansive theory has found use in a variety of cosmological and astrophysical scenarios from inflation \cite{Kaloper:2003yf,Kobayashi:2011nu} to dark energy \cite{DeFelice:2011bh,DeFelice:2010pv,DeFelice:2010nf,Kase:2014yya,Bellini:2014fua} to neutron stars \cite{Cisterna:2015yla,Maselli:2016gxk,Babichev:2016jom,Sakstein:2016oel} and other astrophysical objects \cite{Koyama:2015oma,Saito:2015fza,Sakstein:2015aqx,Sakstein:2015zoa,Sakstein:2015aac,Jain:2015edg,Sakstein:2016ggl,Sakstein:2016lyj}. 

The enormous freedom in constructing models has enabled several examples of black holes (BHs) with scalar hair to be found. These circumvent the no-hair theorem \cite{Bekenstein:1996pn,Sotiriou:2011dz,Faraoni:2013iea} because it was derived assuming only first-order derivatives of the scalar and non-derivative couplings to curvature tensors (a no-hair theorem has been proved for asymptotic BHs in shift-symmetric Horndeski theories \cite{Hui:2012qt} with one loophole \cite{Sotiriou:2013qea,Sotiriou:2014pfa,Benkel:2016rlz}). A comprehensive and systematic review of hairy solutions in Horndeski theories as well as how to construct them can be found in reference \cite{Babichev:2016rlq}. In this work, we are concerned with the specific theory with a non-minimal derivative coupling of the scalar to the graviton
\begin{align}
S=&\int\dd^4 x\sqrt{-g}\left[\frac{m_p^2}{2}\left(R-2\Lambda\right)\right.\nonumber\\&\left.-\frac{1}{2}\left(g^{\mu\nu}-\frac{z}{m_p^2}G^{\mu\nu}\right)\partial_\mu\varphi\partial_\nu\varphi\right],
\label{lag0}
\end{align}
which has been well-studied in the literature\footnote{This specific theory does not include a screening mechanism but passes solar system tests nonetheless since derivatively coupled scalars only source scalar field gradients through their (weak) cosmological dynamics \cite{Sakstein:2014isa,Sakstein:2014aca,Ip:2015qsa,Sakstein:2015jca}.}. Here $m_p$ is the reduced Planck mass, $R$ is Ricci scalar and $G^{\mu\nu}$ is the Einstein tensor. In particular, in the absence of the canonical kinetic term, this theory is a specific example of the \emph{John} class of fab-four theories, which can self-tune away a large cosmological constant \cite{Charmousis:2011bf}. The parameter $z$ is a free coupling constant and $\Lambda$ is a bare cosmological constant. Many cosmological and astrophysical scenarios have been studied in this theory, including inflation \cite{Bruneton:2012zk}, dark matter \cite{Rinaldi:2016oqp}, neutron stars \cite{Cisterna:2016vdx}, etc. Besides, it has been shown that this theory admits hairy BHs that are asymptotically anti-de Sitter (AdS) \cite{Rinaldi:2012vy,Babichev:2013cya,Anabalon:2013oea,Minamitsuji:2013ura}. In the original construction \cite{Rinaldi:2012vy}, the bare cosmological constant was absent and the scalar derivative $\varphi'^2<0$ outside the horizon. Here prime is the derivative with the radial coordinate in the Schwarzschild system. This is problematic since it violates the null energy condition and $\varphi$ is ultimately coupled to matter. Later, \cite{Babichev:2013cya,Anabalon:2013oea,Minamitsuji:2013ura} showed that this pathology could be ameliorated by including a bare cosmological constant. 

Whilst not particularly relevant for cosmology, the study of AdS BHs is especially important for the AdS/CFT correspondence \cite{Maldacena:1997re,Gubser:1998bc,Witten:1998qj}. Large AdS BHs describe (approximate) thermal states of the boundary CFT and it may be the case that AdS BHs in these theories are dual to an interesting strongly coupled three-dimensional gauge theory. Similarly, the decay of a scalar outside the BH---quasi-normal modes (QNMs)---corresponds to perturbations of these states, and contain information about the time-scale for the system to reach equilibrium \cite{Chan:1996yk,Chan:1999sc,Aharony:1999ti,Horowitz:1999jd,Konoplya:2011qq,Zangeneh:2017rhc}. In particular, the QNMs of AdS BHs correspond to poles of the retarded Green's function for the boundary CFT; we refer the reader to \cite{Konoplya:2011qq} and references therein for the applications of this to hydrodynamic systems.

Motivated by this, Minamitsuji has numerically calculated the fundamental QNM for a massless test scalar outside an AdS BH for this theory \cite{Minamitsuji:2014hha}. The purpose of this work is two-fold. First, we extend this calculation to  the higher overtones and non-radial modes. Second, we calculate the QNMs for the case where the scalar is non-minimally coupled to the gravitational scalar $\varphi$; we investigate the lowest-order coupling that preserves the symmetries of $\varphi$ and test scalar. In the former case, we find qualitatively similar behaviour as the fundamental QNMs calculated by reference \cite{Minamitsuji:2014hha}. In the latter case, we find that there is a critical value of the non-minimal coupling below which the effective potential has a different behaviour at asymptotic infinity so that the QNMs are not well-defined. We numerically calculate the QNMs for parameter choices where this is not the case and find that stronger non-minimal couplings increase the oscillation period and decay rate of the QNMs (at fixed BH horizon and derivative coupling constant). 

This paper is organized as follows: in section \ref{sec:AdsBH} we introduce the specific BH studied in this work. The QNMs are calculated and discussed in section \ref{sec:QNM} (for both the minimal and non-minimal coupling) before concluding in section \ref{sec:concs}.

\section{AdS black holes in derivatively-coupled theories}
\label{sec:AdsBH}

The theory defined by the action \eqref{lag0} admits AdS BH solutions of the form \cite{Rinaldi:2012vy,Babichev:2013cya,Anabalon:2013oea,Minamitsuji:2013ura}
\be\label{metric}
ds^2=-F(r)dt^2+\frac{h^2(r)}{F(r)}dr^2+r^2(d\theta^2+\sin^2\theta d\phi^2),
\ee
with 
\begin{align}\label{Fhphi}
F(r)&=1-\frac{2M}{r}+\frac{r^2}{l^2}\nonumber+\frac{(3z-m_p^2l^2)^2}{12zm_p^2l^2}\frac{\arctan(m_pr/\sqrt{z})}{m_pr/\sqrt{z}},\nonumber\\
h(r)&=\frac{z(6m_p^2r^2+m_p^2l^2+3z)}{\sqrt{12z}m_pl(m_p^2r^2+z)},\nonumber\\
\varphi'^2(r)&=\frac{4m_p^6r^2(3z-m_p^2l^2)}{z(m_p^2r^2+z)(3z+m_p^2l^2)}\frac{h^2(r)}{F(r)},
\end{align}

where the AdS length $l$ is related to the coupling constant $z$ and the cosmological constant $\Lambda$ via\footnote{Note that it is not possible to choose a value of $\Lambda$ such that the solution is an asymptotically de Sitter (dS) BH. Such a choice cannot lead to the formation of a cosmological horizon. One could choose $z<0$, but this results in a naked curvature singularity that is not hidden behind a horizon \cite{Minamitsuji:2014hha}. }
\be\label{eq:ADSlength}
l^2=\frac{3z(3m_p^2+z\Lambda)}{m_p^2(m_p^2-z\Lambda)}.
\ee
The BH horizon radius $r_h$ is the only real solution of $F(r_h)=0$, and the Hawking temperature of this horizon is 
\be\label{HawkingT}
T=\frac{F'(r_h)}{4\pi h(r_h)}=\frac{6m_p^2r_h^2+m_p^2l^2+3z}{8\sqrt{3}\pi z m_p l r_h}.
\ee

In order for the solution for $\varphi$ to be real, we need to impose
\be
z\ge m_p^2l^2/3.
\ee
When this lower bound is saturated, one has $\varphi'(r)=0$, $h(r)=1$, and the $\arctan$ term in $F(r)$ vanishes. Therefore, we have an exact Schwarzschild-anti-de Sitter (SAdS) BH\footnote{It is important to note that the cosmological constant for this black hole differs from $\Lambda$ (see Eqn. \eqref{eq:ADSlength}) so that this still represents a non-GR solution. In particular, one would expect metric perturbations to differ from their GR counterparts. In the limit $z\rightarrow0$ the theory reduces to GR, in which case the vacuum solution is an SAdS black hole with AdS length set by $\Lambda$. Since SAdS BHs are solutions of both theories, they are observationally indeistinguishable if one only considers their static, stationary properties, but their different dynamics, such as metric perturbations and interaction with matter, can be used to distingish between the two theories.}. 
In the following, we will first study the QNMs for this case and compare them with known results in the literature  \cite{Horowitz:1999jd,Minamitsuji:2014hha} as a test of our numerical procedure. We will then consider more general values of $z$ where $\varphi'(r)$ is nonzero and the BH deviates from exact SAdS, as well as non-minimal couplings of $\varphi$ to the test scalar. In what follows, we will work in units where $m_p=1$. Furthermore, we will rescale our distances so that $l=1$ i.e. $r$ and $M$ both have units of $l$. 

\section{Quasi-normal Modes}
\label{sec:QNM}
\subsection{Minimal Coupling}

We first consider a test scalar field $\Phi$, minimally coupled to the metric but not to $\varphi$. This is the simplest situation one can envision and we will henceforth refer to it as the \emph{minimally coupled case}. This is in contrast to the case where one has direct couplings between $\varphi$ and $\Phi$, which we will refer to as the \emph{non-minimally coupled case}. The minimally-coupled Lagrangian is
\be\label{lag-coupl}
L_{\rm MC}=-\frac{1}{2}\sqrt{-g}\partial_{\mu}\Phi\partial^{\mu}\Phi,
\ee
and, ignoring the back-reaction of $\Phi$ on the spacetime, the metric is given by Eq. (\ref{metric}) so that the equation of motion of $\Phi$ is 
\be\label{KG}
\Box\Phi=0.
\ee
For our static and spherically symmetric background, one can separate the dependence on coordinates as
\be\label{separation}
\Phi=\frac1{r}\psi(r)Y_j^m(\theta,\phi)e^{-i\omega t}
\ee
where $Y_j^m(\theta,\phi)$ are the usual spherical harmonics with degree $j$ and order $m$. Defining
\be\label{f(r)}
f(r)=F(r)/h(r),
\ee
and introducing the tortoise coordinate $r^*$ given by $dr^*=dr/f(r)$, Eq. (\ref{KG}) can be written in a similar form to the Schr\"{o}dinger equation:
\be\label{schrodinger}
\frac{d^2\psi}{d{r^*}^2}+(\omega^2-V(r))\psi=0,
\ee
where the effective potential is
\be\label{V(r)}
V(r)=\frac{f(r)f'(r)}{r}+j(j+1)\frac{h(r)f(r)}{r^2}.
\ee

\begin{figure}[ht]
\includegraphics[width=3.4in]{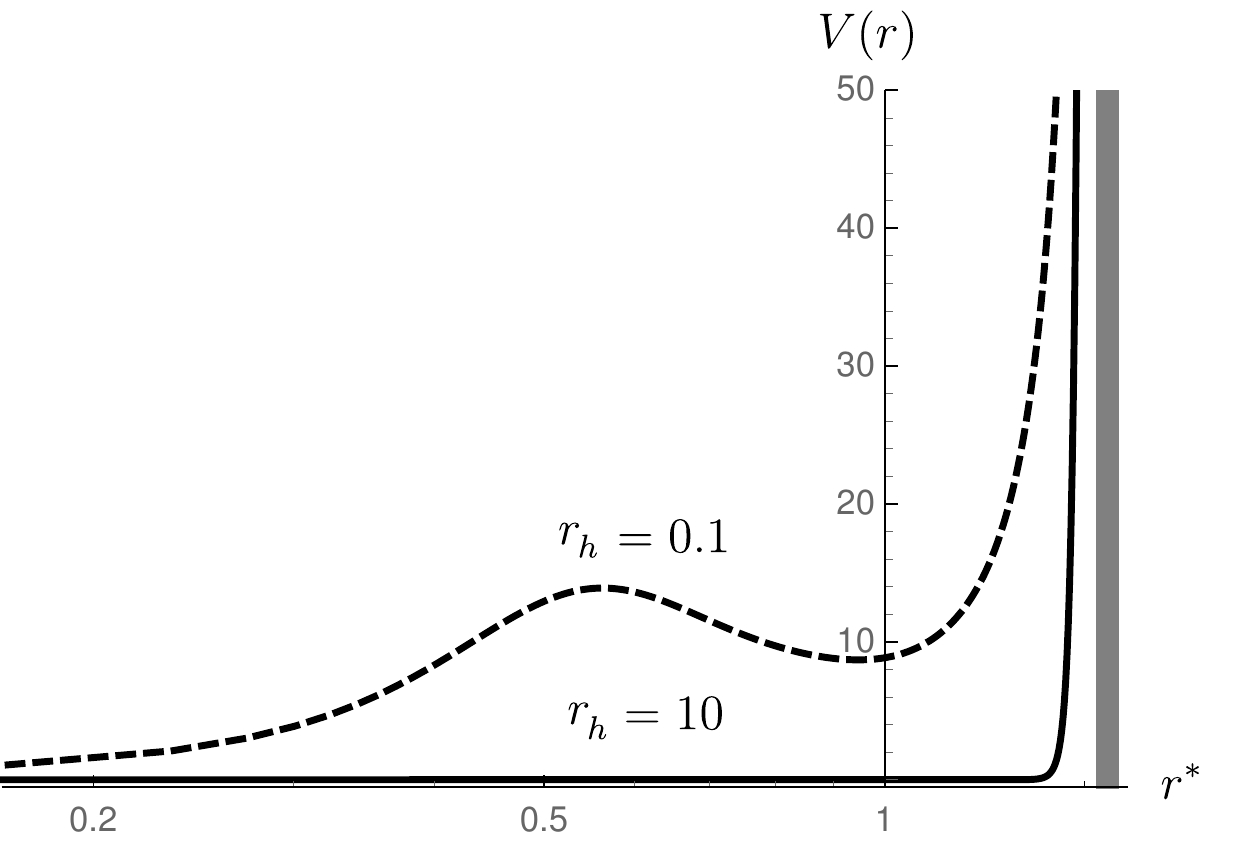}
\caption{Effective potential for a scalar perturbation given in Eq. (\ref{V(r)}). The tortoise coordinate $r^*$ takes values from -$\infty$ to $\pi/2$. $z=1/3$ and $j=0$. }\label{fig-V-r_ads}
\end{figure}

We plot $V(r)$ for large and small BHs for the SAdS case $z=1/3$ in Fig. \ref{fig-V-r_ads}. Consider perturbations outside the BH, i.e. $r_h<r<\infty$. From Eqs. (\ref{Fhphi}) and (\ref{f(r)}), we see that $f\approx4\pi T(r-r_h)$ as $r\rightarrow r_h$, and $f\approx C_1 r^2$ as $r\rightarrow\infty$. $C_1$ is a positive constant. From the definition of the tortoise coordinate, we find
\bea
r^*&\approx&\frac1{4\pi T}\ln(r-r_h),\quad r\rightarrow r_h;\nonumber\\
r^*&\approx&C_2-\frac1{C_1 r},\quad r\rightarrow\infty.
\eea
Here $C_2$ is an integration constant which can be freely chosen. We set it to $\pi/2$. Clearly, $r^*$ tends to $-\infty$ as $r$ approaches $r_h$ so this coordinate takes values in the range $-\infty<r^*<\pi/2$. It is evident that, for any value of $z$, $V(r)$ vanishes as $r^*$ goes to $-\infty$ (as $r$ approaches $r_h$), and $V(r)$ diverges as $r^*$ goes to its upper bound $\frac{\pi}{2}$ (as $r$ goes to $\infty$), as shown in Fig. \ref{fig-V-r_ads}. 

The QNMs are then naturally defined as the complex values of $\omega=\omega_{ST}$, so that the solution of Eq. (\ref{schrodinger}) has the following asymptotic form,
\bea\label{bc}
\psi&\sim& e^{-i\omega r^*},\, r\rightarrow r_h;\nonumber\\
\psi&\rightarrow&0,\, r\rightarrow \infty.\label{eq:BCS}
\eea
We apply the numerical approach proposed in \cite{Horowitz:1999jd} to solve for the QNMs; the details of this method are outlined in appendix \ref{sec:numerical}. Note that for SAdS BHs in GR, the imaginary parts of the QNM are always negative ($\Im(\omega_{GR})<0$) \cite{Horowitz:1999jd}. This implies that the modes always decay. The same is true for all asymptotically AdS BHs in the theory we consider here, i.e. $\Im(\omega_{ST})<0$; we refer the reader to reference \cite{Horowitz:1999jd} for a formal proof. 

\subsubsection{Quasi-normal Modes for $z=1/3$}

As discussed above, when $z=1/3$ we have $h(r)=1$ and $f(r)=F(r)=1-\frac{2M}{r}+r^2$ so that the metric has precisely SAdS form (note that $\varphi'(r)=0$). The QNMs will then be those of the SAdS BH, even if there is a finite coupling between $\Phi$ and $\partial\varphi$. We begin by studying the QNMs for the radial perturbations, i.e. $j=0$ modes, for $r_h$ between $10$ and $1/4$, for the principal QNM and the first two overtones. As $r_h$ becomes smaller, a larger order of expansion is needed to get precise solutions. For example, 50 orders are sufficient for $r_h=10$, while 450 orders are considered for $r_h=1/4$ in order to be precise to 3 decimal places. 
Our results for typical values of $r_h$ are shown in Tab. (\ref{tab-w012}) in Appendix B and are plotted in Fig. \ref{fig-QNM_SAdS}.

\begin{figure}[ht]
  \begin{tabular}{@{}c@{}}
    \includegraphics[width=.9\linewidth]{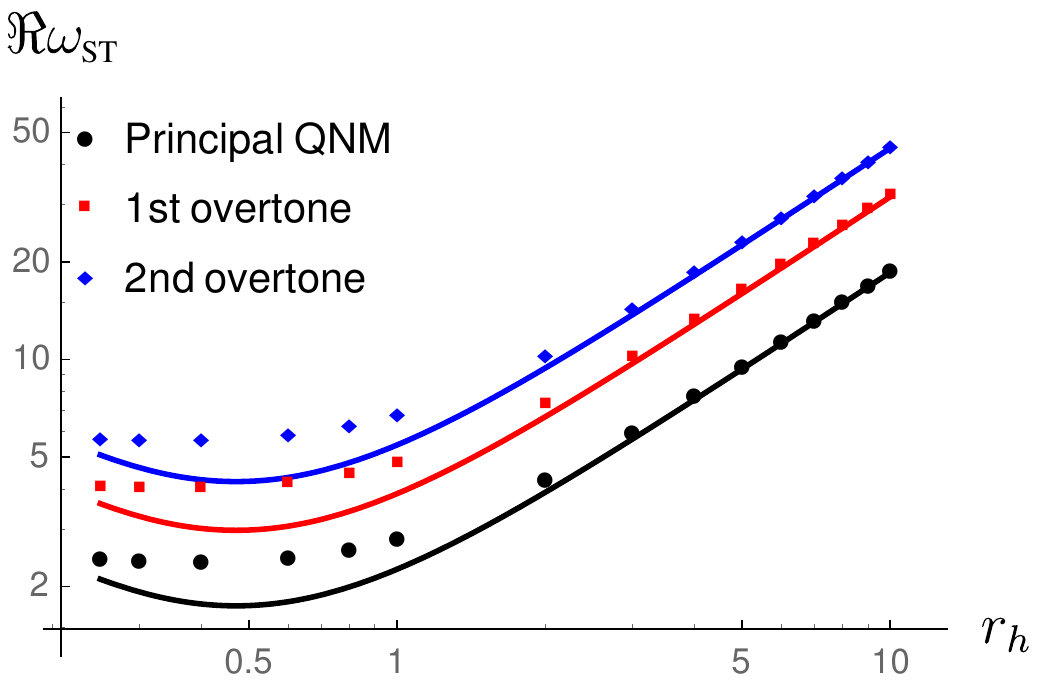} \\[\abovecaptionskip]
    \small (a) $\Re(\omega_{ST})$.
  \end{tabular}

  \vspace{\floatsep}

  \begin{tabular}{@{}c@{}}
    \includegraphics[width=.9\linewidth]{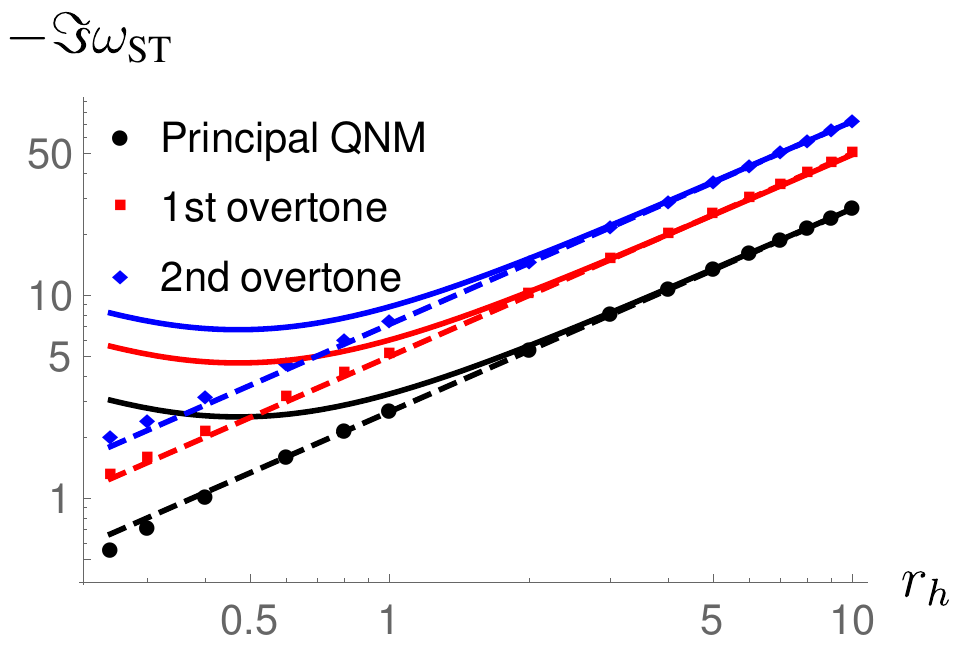} \\[\abovecaptionskip]
    \small (b) $-\Im(\omega_{ST})$.
  \end{tabular}
  \caption{Real (a) and imaginary (b) parts of the QNMs, ($\omega_{ST}$), as a function of the BH horizon radius $r_h$. Here $z=1/3$. The data points are the principal QNM (black circles), and the first (red squares) and second (blue diamonds) overtones. The solid continuous curves are $(7.75-11.16i)T$ (principal), $(13.24-20.59i)T$ (first overtone) and $(18.70-30.02i)T$ (second overtone), where $T$ is the Hawking temperature of the BH horizon. The dashed continuous curves in (b) are $2.66r_h$ (principal), $4.98r_h$ (first overtone) and $7.18r_h$ (second overtone). The units are chosen by setting $m_p=1$ and $l=1$.\label{fig-QNM_SAdS}}
\end{figure}

For large BHs, the relations $\omega_{ST}^{(0)}=(7.75-11.16i)T$, $\omega_{ST}^{(1)}=(13.24-20.59i)T$, and $\omega_{ST}^{(2)}=(18.70-30.02i)T$, as found for SAdS BHs in \cite{Horowitz:1999jd}, hold. Here $\omega_{ST}^{(0)},~\omega_{ST}^{(1)},~\omega_{ST}^{(2)}$ are the principal QNM and the first and second overtones respectively, and $T$ is the Hawking temperature of the BH horizon given in Eq. (\ref{HawkingT}). As $r_h$ decreases, the above linear relations no longer hold. For intermediate-size BHs, a linear relation between $\Im(\omega_{ST})$ and $r_h$ holds, i.e. $\Im(\omega_{ST}^{(0)})=-2.66r_h$, $\Im(\omega_{ST}^{(1)})=-4.98r_h$ and $\Im(\omega_{ST}^{(2)})=-7.18r_h$. This relation breaks down for smaller BHs. As found by \cite{Konoplya:2002zu}, the QNMs of SAdS BHs approach those of a pure AdS space as the hole becomes very small.

Next, we consider non-radial perturbations, i.e. $j>0$. It is necessary to go to larger orders in the expansion in order to get convergent results for larger values of $j$. We were able to calculate the principal QNM for $j$ up to 30, for $r_h$ down to 4. Typical results are listed in Tab. (\ref{tab-wl30}) in Appendix B, and these are plotted in Fig. \ref{fig-w-j}. As seen, both of the real and imaginary parts of QNMs increase with $j$, with the change becoming less significant for larger BHs. Our results are consistent with those of \cite{Horowitz:1999jd}, who have studied SAdS BHs previously.

\begin{figure}[ht]
\includegraphics[width=3.4in]{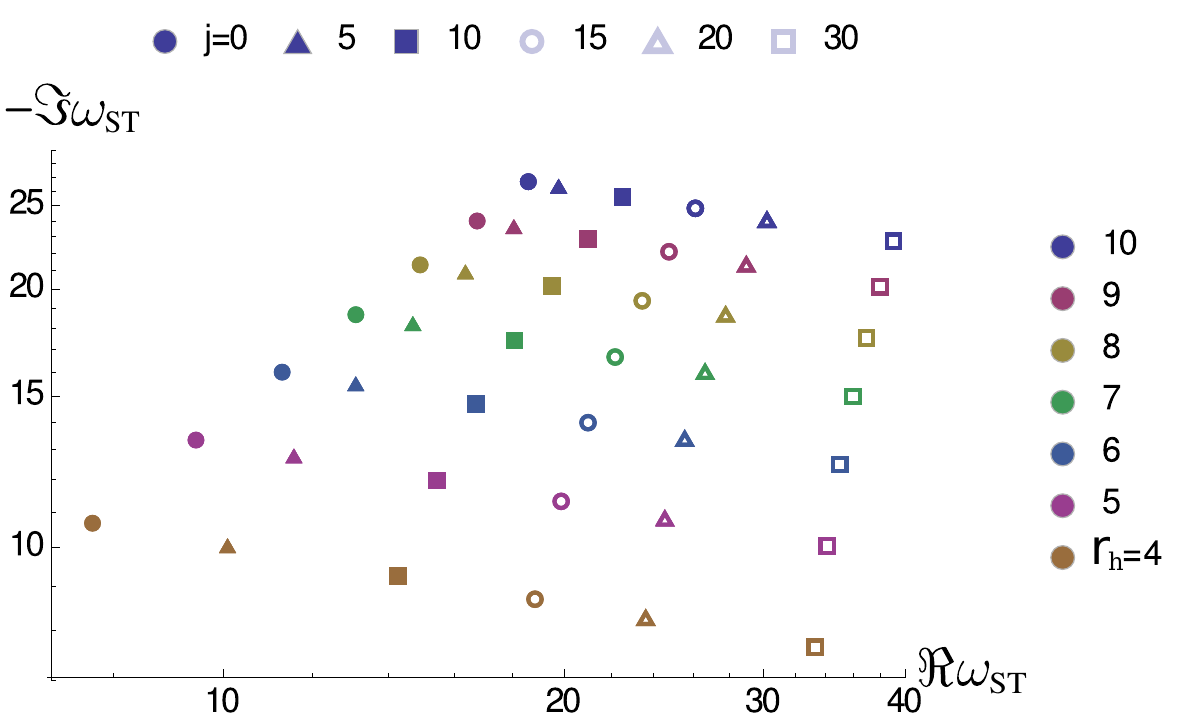}
\caption{Contour plot of the principal QNMs for different values of $j$ and $r_h$. Different $j$'s are denoted by different marker shapes. Different $r_h$'s are represented by different marker colors. Here $z=1/3$. \label{fig-w-j}}
\end{figure}

\subsubsection{Quasi-normal Modes for $z>1/3$}

When $z>1/3$, two factors contribute to the change in QNMs: the spacetime (\ref{metric}) deviates from SAdS, and there is a potential coupling between the test field and $\varphi$. In this section, we consider the former, the latter is the topic of the next section. As before, we begin by considering radial perturbations, $j=0$. Examining Eq. (\ref{schrodinger}) (using $V(r)$ as given in Eq. (\ref{V(r)})), the differences from SAdS are due to the different form of $f(r)$ in Eq. (\ref{f(r)}). For BHs with $r_h>>\sqrt{z}$ we have,
\begin{align}
\arctan(r/\sqrt{z})&\approx\pi/2,\quad\textrm{and}\\
h(r)&\approx\sqrt{3z},
\end{align}
and therefore $f(r)$ differs from the SAdS form by a constant factor $\sqrt{\frac{1}{3z}}$. The potential $V(r)$ then differs from the SAdS potential by $\frac{1}{3z}$. As a result,
\be\label{eq:scalingwithz}
\omega_{ST}\approx\sqrt{\frac{1}{3z}}\omega_{GR},
\ee
where $\omega_{GR}$ is the corresponding QNM for an SAdS BH (solution in GR) with the same horizon $r_h$.

\begin{figure}[h]
  \begin{tabular}{@{}c@{}}
    \includegraphics[width=.9\linewidth]{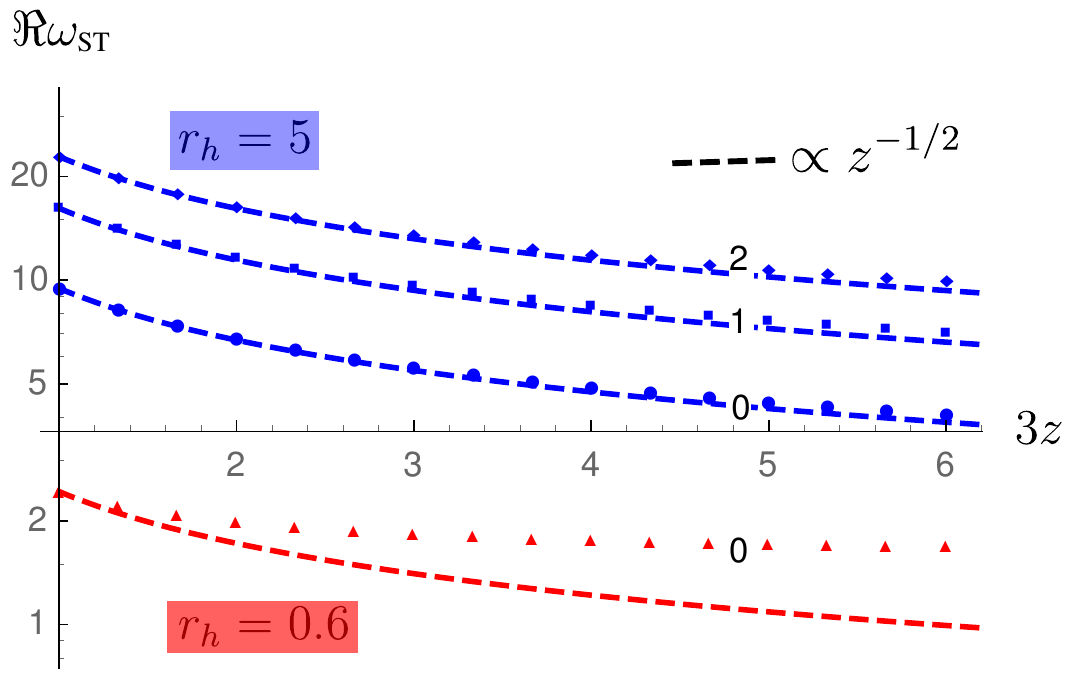} \\[\abovecaptionskip]
    \small (a) $\Re(\omega_{ST})$.
  \end{tabular}
  \vspace{\floatsep}
  \begin{tabular}{@{}c@{}}
    \includegraphics[width=.9\linewidth]{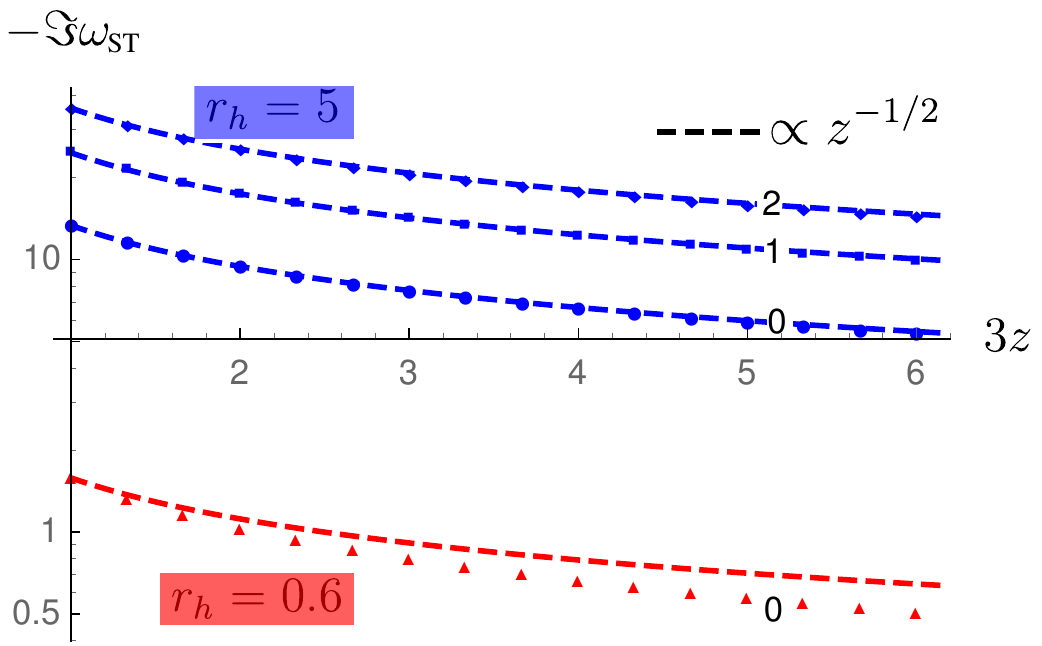} \\[\abovecaptionskip]
    \small (b) $-\Im(\omega_{ST})$.
  \end{tabular}
 \caption{The real (a) and imaginary (b) parts of the QNMs as a function of the constant $z$. The dashed lines are given by Eq.\eqref{eq:scalingwithz} with the corresponding BH horizon radius and QNM order. Different $r_h$'s are represented by different colors. The integers between data points are the order of QNMs: "0" for principal mode, "1" for the first overtone and "2" for the second overtones.  \label{w-z}}
\end{figure}

We plot the above relation as dashed lines in Fig. \ref{w-z} for the principle QNM as well as the first and second overtones, and compare them with our numerically computed QNMs (only the principal mode is shown for $r_h=0.6$ since it is sufficient to illustrate that the relation breaks down for small BHs). As expected, our data points follow these lines very well for large BHs but the deviation is significant for small BHs, i.e. $r_h<\sqrt{z}$. This can be seen in the figure as small deviations at large $z$ for the case $r_h=5$.

In general, $\Re(\omega_{ST})$ and $-\Im(\omega_{ST})$ decrease with increasing $z$. Physically, this means the dominant perturbation oscillates with a longer period and decays more slowly as $z$ increases. From an AdS/CFT correspondence point of view, this means it takes a longer time to reach equilibrium. For large BHs, this change follows the $z^{-1/2}$ trend as discussed above. For small BHs, $\Re(\omega_{ST})$ changes more slowly while $-\Im(\omega_{ST})$ changes more rapidly with $z$.

\begin{figure}[h]
  \includegraphics[width=.9\linewidth]{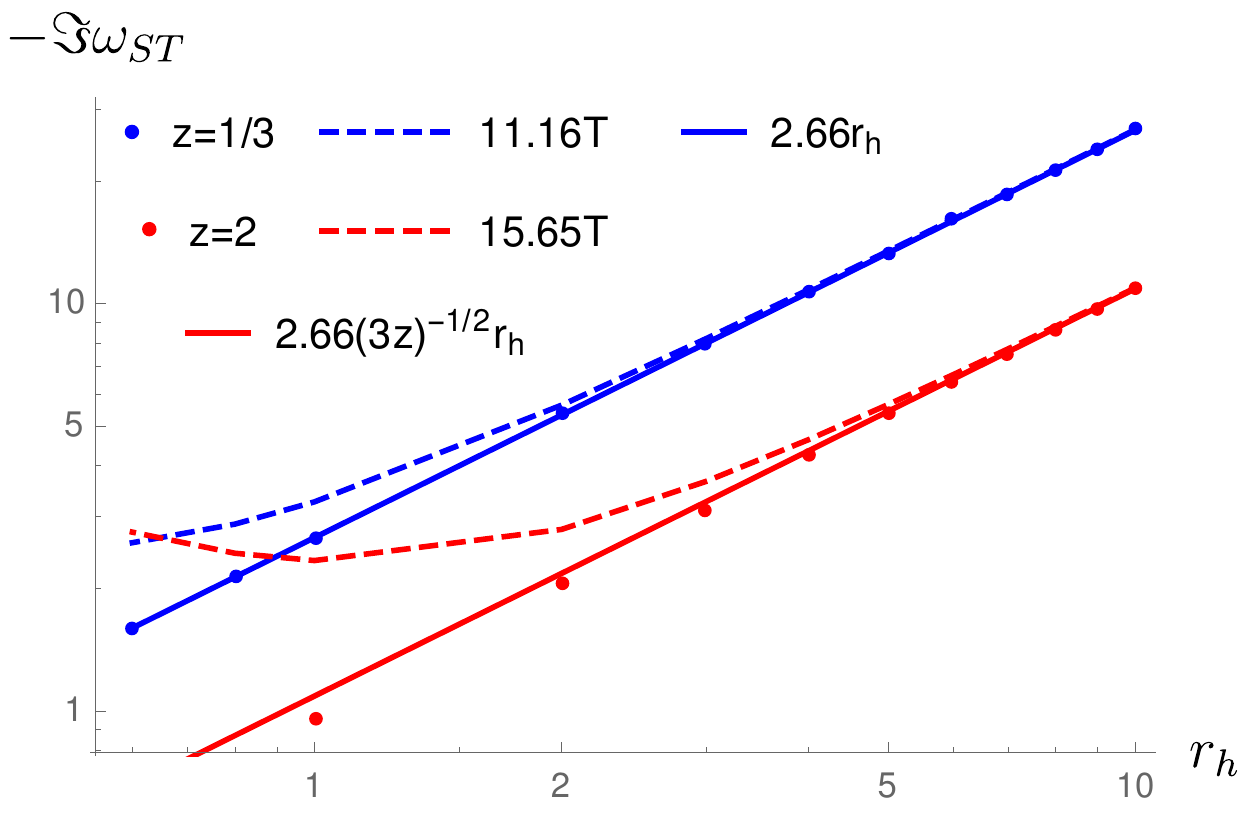} \\[\abovecaptionskip]
 \caption{Imaginary part of QNMs as a function of the BH horizon radius, for $z=1/3$ (blue) and $z=2$ (Red). The filled circles are data from our calculation, and the solid/dashed curved are the analytic functions shown in the legend. \label{fig-wI-rh_z}}
\end{figure}

We also plot the imaginary part of the QNMs as a function of $r_h$ for $z=2$ in Fig. \ref{fig-wI-rh_z}. As discussed in the previous subsection, $\Im(\omega_{ST})$ is proportional to $T$ for large BHs, and to $r_h$ for intermediate-size BHs. As seen here, the linear relation with $T$ still holds for large BHs for $z>1/3$, with different proportionality. And the linear relation with $r_h$ also holds for intermediate-size BHs, with the proportionality reduced by a factor of $(3z)^{1/2}$. Note that these linear relations break down at larger values of $r_h$ than the SAdS BH relations. As discussed by \cite{Horowitz:1999jd}, $\Re(\omega_{ST})$ never scales as $r_h$ no matter the value of $z$.

\begin{figure}[ht]
\includegraphics[width=3.4in]{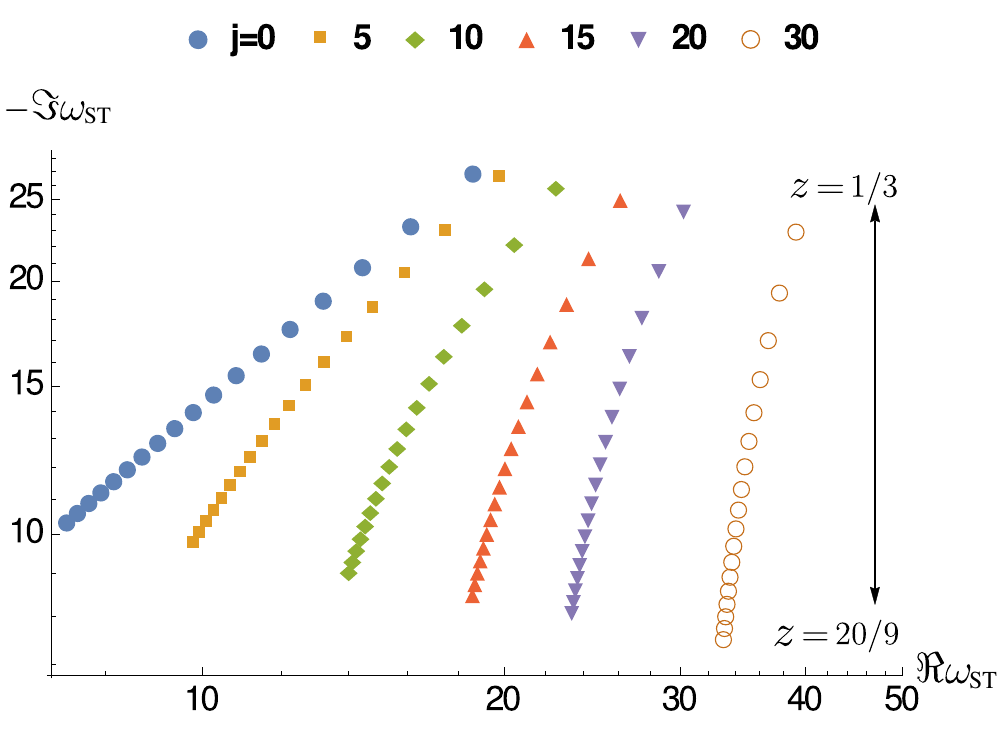}
\caption{Contour plot of the principal QNMs for different $j$'s and $z$'s for $r_h=10$. Different $j$'s are denoted by different marker shapes shown in the figure. For each $j$, $z$ takes values uniformly from 1/3 to 20/9, from top to bottom. \label{fig-w-j-z}}
\end{figure}

Next, we compute the non-radial QNMs for different values of $z$ and plot them in Fig. \ref{fig-w-j-z}. We choose $r_h=10$.\footnote{The QNMs for different values of $r_h$ (provided $r_h>1$ and $r_h>\sqrt{z}$) have similar dependencies on $j$ and $z$.} This constitutes one of the new results of this work. For each value of $z$, higher-$j$ order QNMs have both larger real and imaginary parts. And the change of QNMs with $j$ becomes more significant for larger values of $z$. 

Comparing Fig. \ref{fig-w-j-z} with Fig. \ref{fig-w-j} shows that increasing $z$ has a similar effect to decreasing $r_h$. The reason for this can be seen from the the metric. For $r_h\gg 1$ and $r_h\gg \sqrt{z}$, the metric functions (Eq. (\ref{Fhphi})) show that the metric is approximately SAdS with
\be
r_h^3\approx 2M-\frac{\pi(3z-1)^2}{24\sqrt{z}}.
\ee
This clearly shows that, for large BHs and not too large $z$, increasing $z$ would reduce $r_h$ while keeping the approximate SAdS form and AdS  radius $l$ of the BH fixed.\footnote{Of course, as discussed in Eq. (\ref{eq:scalingwithz}), increasing $z$ causes a decrease of $\Re(\omega_{ST})$ and $-\Im(\omega_{ST})$, due to the first term in $V(r)$ (Eq. (\ref{V(r)})). The second term in $V(r)$ is exactly the same as the SAdS counterpart for large BHs. As $j$ becomes larger, this term is more important and increasing $z$ has an effect more similar to that of decreasing $r_h$.}

\subsection{Non-minimal Coupling}

\begin{figure}[ht]
  \includegraphics[width=.9\linewidth]{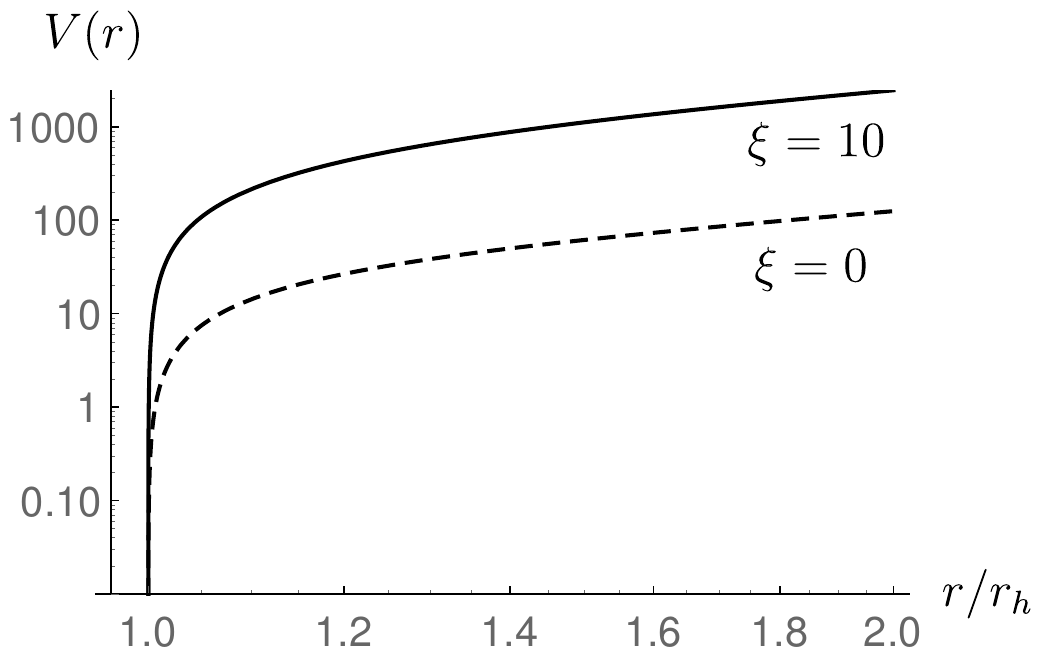} 
  \caption{The effective potential for a non-minimally coupled scalar model (Eq. (\ref{Vr-xi})) with different values of the coupling constant $\xi$, for $r_h=10$, $j=0$ and $z=2/3$. The critical value of $\xi$ is $\xi_c=-1/2$.   \label{fig-V-r_xi}}
\end{figure}

In this subsection we consider a coupling between the test scalar and the gravitational scalar $\varphi$. The simplest form of coupling preserving the shift symmetry of the field $\varphi$, $\varphi\rightarrow\varphi+c$ with $c$ constant, and the reflection symmetry of $\Phi$ ($\Phi\rightarrow -\Phi$) and $\varphi$ ($\varphi\rightarrow -\varphi$) is
\be\label{lagrangian-xi}
L_{\rm NMC}=\sqrt{-g}\left(-\frac12\partial_{\mu}\Phi\partial^{\mu}\Phi-\frac{\xi}{2 m_p^2}\Phi^2\partial_{\mu}\varphi\partial^{\mu}\varphi\right),
\ee
where $\xi$ is a dimensionless coupling constant. The equation of motion for $\Phi$ is now modified to
\be
\Box\Phi-\frac{\xi}{m_p^2}(\partial_{\mu}\varphi\partial^{\mu}\varphi)\Phi=0,
\ee
which still reduces to the form of a Schr\"{o}dinger-like equation for $\psi$ when written in terms of the tortoise coordinate, but with the effective potential 
\be
V(r)=\frac{ff'}{r}+j(j+1)\frac{hf}{r^2}+\frac{\xi}{m_p^2}f^2\varphi'^2(r),\label{Vr-xi}
\ee
where $\varphi'(r)$ is given by Eq. \eqref{Fhphi}. The first and third terms both vary as $r^2$ when $r\rightarrow\infty$, and therefore there is a critical value of $\xi$ above which $V(r)\rightarrow +\infty$ as $r\rightarrow\infty$, but below which $V(r)\rightarrow -\infty$ as $r\rightarrow\infty$. This critical value is 
\be\label{xi_c}
\xi_c=-\frac{3z+1}{6(3z-1)}.
\ee
At $\xi=\xi_c$, $V(r)$ is dominated by the second term and becomes constant as $r\rightarrow\infty$. For $j=0$, $V(r)\sim 1/r\rightarrow 0$ as $r\rightarrow\infty$. For $\xi\le\xi_c$, our boundary conditions (Eq. (\ref{bc})) in the definition of QNMs no longer apply. For this reason, we will only consider values of $\xi$ above $\xi_c$. The effective potential is plotted in Fig. \ref{fig-V-r_xi} for this case for different values of $\xi$.

\begin{figure}[h]
\includegraphics[width=3.4in]{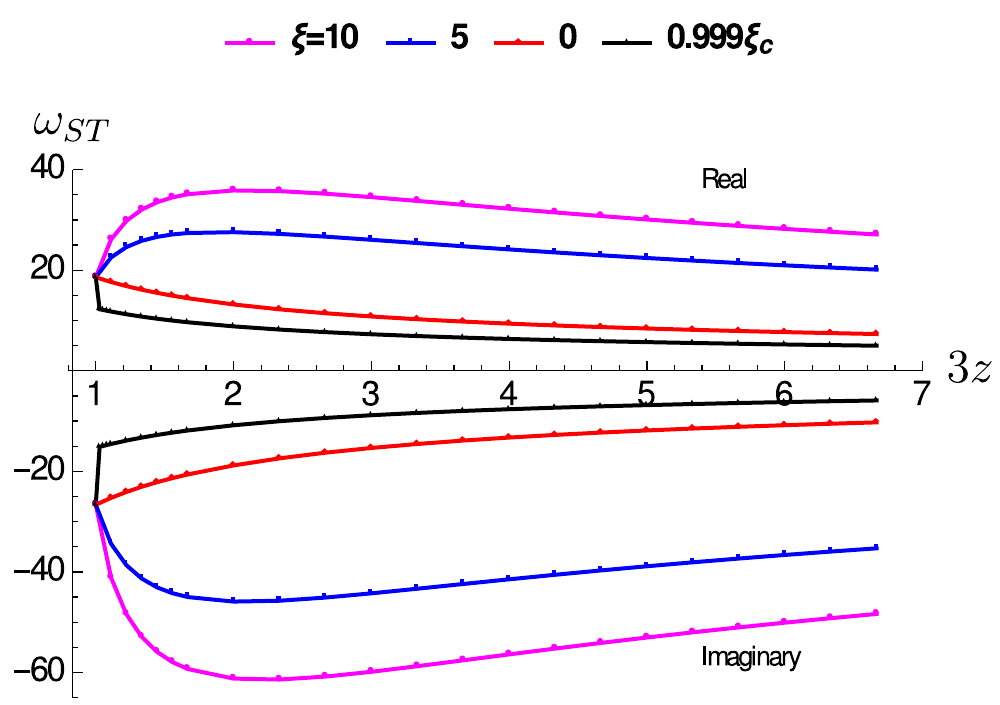}
\caption{Real part (upper half plane) and imaginary part (lower half plane) of the principal QNMs as a function of $z$ for different values of the coupling constant $\xi$ (defined in Eq. (\ref{lagrangian-xi})) given in the figure. Here $r_h=10$ and $j=0$. Neighboring data points are joined by line segments for better illustration. \label{fig-w-xi}}
\end{figure}

The principal QNMs are plotted in Fig. \ref{fig-w-xi}, for $j=0$ and $r_h=10$. As seen, both $\Re(\omega_{ST})$ and $-\Im(\omega_{ST})$ increase as $\xi$ increases. For low values of $\xi$, $\Re(\omega_{ST})$ and $-\Im(\omega_{ST})$ decrease as $z$ increases but for large $\xi$, $\Re(\omega_{ST})$ and $-\Im(\omega_{ST})$ increase with $z$ for small $z$, and then decrease for large $z$. Recall that when $z=1/3$ the solution is exactly SAdS with $\varphi'=0$ so that the non-minimal coupling is not relevant and the QNMs converge to the same value whatever the value of $\xi$. At $z=1/3$, $\xi_c$ becomes $-\infty$ (Eq. (\ref{xi_c})). Therefore, for $\xi$ very close to $\xi_c$, the QNMs have a very steep change with $z$ close to $z=1/3$. This is clearly seen in Fig. \ref{fig-w-xi}.

For a fixed value of $z$, the real part of QNM increases while the imaginary part becomes more negative with increasing $\xi$. Physically this means the dominant perturbation will have a shorter oscillation period and will decay more rapidly. In terms of AdS/CFT, if such BHs have a CFT dual, this means the equilibrium state is reached faster.

\section{Conclusions}
\label{sec:concs}

In this work we have studied the quasi-normal modes of asymptotically Anti-de Sitter black holes that are analytic solutions of a class of shift-symmetric Horndeski theories where a gravitational scalar $\varphi$ is derivatively coupled to the Einstein tensor. We have calculated the QNMs numerically for a massless test scalar both minimally coupled to the metric, and non-minimally coupled to $\varphi$. 

In the case of minimal coupling, we have calculated the principal radial as well as the first two overtones for parameter choices that give exact SAdS solutions. A linear relation between the (complex) frequency $\omega_{ST}$ and the Hawking temperature was observed in all cases for large black holes, confirming known analytic expectations. We also calculated the principal mode for non-radial perturbations and found that increasing the horizon radius increases the real part of $\omega_{ST}$ and makes the imaginary part more negative. 

Moving away from exact Schwarzschild-Anti-de Sitter black holes, we calculated the principal radial mode and first two overtones and found that increasing the coupling $z$ of $\varphi$ to gravity decreases the real part and makes the imaginary part of $\omega_{ST}$ less negative at fixed black hole radius. We predict and numerically confirm the relation $\omega_{ST}\propto z^{-1/2}$ for large black holes. In the context of the AdS/CFT correspondence, the dual theory exhibits perturbations from the thermal state that decay more slowly. We also calculated the principal non-radial mode and found that, for a black hole with a fixed radius, stronger couplings to gravity have a similar effect to decreasing the horizon radius in the case of Schwarzschild-Anti-de Sitter black holes.

Finally, we considered, for the first time, a non-minimal coupling between $\varphi$ and the test scalar $\Phi$; we chose the lowest-order operator that respects the symmetries of both fields. We found that there is a critical value of the dimensionless coupling constant $\xi$ below which the structure of the effective potential for perturbations changes so that $\lim_{r\rightarrow\infty}V(r)=-\infty$ and the QNMs satisfying the usual boundary conditions (Eq. (\ref{eq:BCS})) are not well-defined. We numerically calculated the principal radial mode for values of $\xi$ larger than this critical value and found that stronger non-minimal couplings increase the real parts and decrease (make more negative) the imaginary parts of the frequency i.e. they give rise to QNMs that oscillate with a shorter period and decay faster than the equivalent (same $z$) minimally coupled models.

\begin{acknowledgments}
We are grateful to Eugeny Babichev and Onkar Parrikar for useful discussions. D.S. and R.D. were partially supported by the US National Science Foundation, under Grant No. PHY-1417317.  JS is supported by funds provided to the Center for Particle Cosmology by the University of Pennsylvania.
\end{acknowledgments}

\bibliography{ref}

\begin{thebibliography}{66}%
\makeatletter
\providecommand \@ifxundefined [1]{%
 \@ifx{#1\undefined}
}%
\providecommand \@ifnum [1]{%
 \ifnum #1\expandafter \@firstoftwo
 \else \expandafter \@secondoftwo
 \fi
}%
\providecommand \@ifx [1]{%
 \ifx #1\expandafter \@firstoftwo
 \else \expandafter \@secondoftwo
 \fi
}%
\providecommand \natexlab [1]{#1}%
\providecommand \enquote  [1]{``#1''}%
\providecommand \bibnamefont  [1]{#1}%
\providecommand \bibfnamefont [1]{#1}%
\providecommand \citenamefont [1]{#1}%
\providecommand \href@noop [0]{\@secondoftwo}%
\providecommand \href [0]{\begingroup \@sanitize@url \@href}%
\providecommand \@href[1]{\@@startlink{#1}\@@href}%
\providecommand \@@href[1]{\endgroup#1\@@endlink}%
\providecommand \@sanitize@url [0]{\catcode `\\12\catcode `\$12\catcode
  `\&12\catcode `\#12\catcode `\^12\catcode `\_12\catcode `\%12\relax}%
\providecommand \@@startlink[1]{}%
\providecommand \@@endlink[0]{}%
\providecommand \url  [0]{\begingroup\@sanitize@url \@url }%
\providecommand \@url [1]{\endgroup\@href {#1}{\urlprefix }}%
\providecommand \urlprefix  [0]{URL }%
\providecommand \Eprint [0]{\href }%
\providecommand \doibase [0]{http://dx.doi.org/}%
\providecommand \selectlanguage [0]{\@gobble}%
\providecommand \bibinfo  [0]{\@secondoftwo}%
\providecommand \bibfield  [0]{\@secondoftwo}%
\providecommand \translation [1]{[#1]}%
\providecommand \BibitemOpen [0]{}%
\providecommand \bibitemStop [0]{}%
\providecommand \bibitemNoStop [0]{.\EOS\space}%
\providecommand \EOS [0]{\spacefactor3000\relax}%
\providecommand \BibitemShut  [1]{\csname bibitem#1\endcsname}%
\let\auto@bib@innerbib\@empty
\bibitem [{\citenamefont {Copeland}\ \emph {et~al.}(2006)\citenamefont
  {Copeland}, \citenamefont {Sami},\ and\ \citenamefont
  {Tsujikawa}}]{Copeland:2006wr}%
  \BibitemOpen
  \bibfield  {author} {\bibinfo {author} {\bibfnamefont {E.~J.}\ \bibnamefont
  {Copeland}}, \bibinfo {author} {\bibfnamefont {M.}~\bibnamefont {Sami}}, \
  and\ \bibinfo {author} {\bibfnamefont {S.}~\bibnamefont {Tsujikawa}},\ }\href
  {\doibase 10.1142/S021827180600942X} {\bibfield  {journal} {\bibinfo
  {journal} {Int.J.Mod.Phys.}\ }\textbf {\bibinfo {volume} {D15}},\ \bibinfo
  {pages} {1753} (\bibinfo {year} {2006})},\ \Eprint
  {http://arxiv.org/abs/hep-th/0603057} {arXiv:hep-th/0603057 [hep-th]}
  \BibitemShut {NoStop}%
\bibitem [{\citenamefont {Clifton}\ \emph {et~al.}(2012)\citenamefont
  {Clifton}, \citenamefont {Ferreira}, \citenamefont {Padilla},\ and\
  \citenamefont {Skordis}}]{Clifton:2011jh}%
  \BibitemOpen
  \bibfield  {author} {\bibinfo {author} {\bibfnamefont {T.}~\bibnamefont
  {Clifton}}, \bibinfo {author} {\bibfnamefont {P.~G.}\ \bibnamefont
  {Ferreira}}, \bibinfo {author} {\bibfnamefont {A.}~\bibnamefont {Padilla}}, \
  and\ \bibinfo {author} {\bibfnamefont {C.}~\bibnamefont {Skordis}},\ }\href
  {\doibase 10.1016/j.physrep.2012.01.001} {\bibfield  {journal} {\bibinfo
  {journal} {Phys.Rept.}\ }\textbf {\bibinfo {volume} {513}},\ \bibinfo {pages}
  {1} (\bibinfo {year} {2012})},\ \Eprint {http://arxiv.org/abs/1106.2476}
  {arXiv:1106.2476 [astro-ph.CO]} \BibitemShut {NoStop}%
\bibitem [{\citenamefont {Joyce}\ \emph {et~al.}(2014)\citenamefont {Joyce},
  \citenamefont {Jain}, \citenamefont {Khoury},\ and\ \citenamefont
  {Trodden}}]{Joyce:2014kja}%
  \BibitemOpen
  \bibfield  {author} {\bibinfo {author} {\bibfnamefont {A.}~\bibnamefont
  {Joyce}}, \bibinfo {author} {\bibfnamefont {B.}~\bibnamefont {Jain}},
  \bibinfo {author} {\bibfnamefont {J.}~\bibnamefont {Khoury}}, \ and\ \bibinfo
  {author} {\bibfnamefont {M.}~\bibnamefont {Trodden}},\ }\href@noop {} {\
  (\bibinfo {year} {2014})},\ \Eprint {http://arxiv.org/abs/1407.0059}
  {arXiv:1407.0059 [astro-ph.CO]} \BibitemShut {NoStop}%
\bibitem [{\citenamefont {Berti}\ \emph {et~al.}(2015)\citenamefont {Berti}
  \emph {et~al.}}]{Berti:2015itd}%
  \BibitemOpen
  \bibfield  {author} {\bibinfo {author} {\bibfnamefont {E.}~\bibnamefont
  {Berti}} \emph {et~al.},\ }\href {\doibase 10.1088/0264-9381/32/24/243001}
  {\bibfield  {journal} {\bibinfo  {journal} {Class. Quant. Grav.}\ }\textbf
  {\bibinfo {volume} {32}},\ \bibinfo {pages} {243001} (\bibinfo {year}
  {2015})},\ \Eprint {http://arxiv.org/abs/1501.07274} {arXiv:1501.07274
  [gr-qc]} \BibitemShut {NoStop}%
\bibitem [{\citenamefont {Koyama}(2015)}]{Koyama:2015vza}%
  \BibitemOpen
  \bibfield  {author} {\bibinfo {author} {\bibfnamefont {K.}~\bibnamefont
  {Koyama}},\ }\href@noop {} {\  (\bibinfo {year} {2015})},\ \Eprint
  {http://arxiv.org/abs/1504.04623} {arXiv:1504.04623 [astro-ph.CO]}
  \BibitemShut {NoStop}%
\bibitem [{\citenamefont {Burrage}\ and\ \citenamefont
  {Sakstein}(2016)}]{Burrage:2016bwy}%
  \BibitemOpen
  \bibfield  {author} {\bibinfo {author} {\bibfnamefont {C.}~\bibnamefont
  {Burrage}}\ and\ \bibinfo {author} {\bibfnamefont {J.}~\bibnamefont
  {Sakstein}},\ }\href {\doibase 10.1088/1475-7516/2016/11/045} {\bibfield
  {journal} {\bibinfo  {journal} {JCAP}\ }\textbf {\bibinfo {volume} {1611}},\
  \bibinfo {pages} {045} (\bibinfo {year} {2016})},\ \Eprint
  {http://arxiv.org/abs/1609.01192} {arXiv:1609.01192 [astro-ph.CO]}
  \BibitemShut {NoStop}%
\bibitem [{\citenamefont {Vainshtein}(1972)}]{Vainshtein:1972sx}%
  \BibitemOpen
  \bibfield  {author} {\bibinfo {author} {\bibfnamefont {A.}~\bibnamefont
  {Vainshtein}},\ }\href {\doibase 10.1016/0370-2693(72)90147-5} {\bibfield
  {journal} {\bibinfo  {journal} {Phys.Lett.}\ }\textbf {\bibinfo {volume}
  {B39}},\ \bibinfo {pages} {393} (\bibinfo {year} {1972})}\BibitemShut
  {NoStop}%
\bibitem [{\citenamefont {Nicolis}\ \emph {et~al.}(2009)\citenamefont
  {Nicolis}, \citenamefont {Rattazzi},\ and\ \citenamefont
  {Trincherini}}]{Nicolis:2008in}%
  \BibitemOpen
  \bibfield  {author} {\bibinfo {author} {\bibfnamefont {A.}~\bibnamefont
  {Nicolis}}, \bibinfo {author} {\bibfnamefont {R.}~\bibnamefont {Rattazzi}}, \
  and\ \bibinfo {author} {\bibfnamefont {E.}~\bibnamefont {Trincherini}},\
  }\href {\doibase 10.1103/PhysRevD.79.064036} {\bibfield  {journal} {\bibinfo
  {journal} {Phys.Rev.}\ }\textbf {\bibinfo {volume} {D79}},\ \bibinfo {pages}
  {064036} (\bibinfo {year} {2009})},\ \Eprint {http://arxiv.org/abs/0811.2197}
  {arXiv:0811.2197 [hep-th]} \BibitemShut {NoStop}%
\bibitem [{\citenamefont {Kaloper}\ \emph {et~al.}(2011)\citenamefont
  {Kaloper}, \citenamefont {Padilla},\ and\ \citenamefont
  {Tanahashi}}]{Kaloper:2011qc}%
  \BibitemOpen
  \bibfield  {author} {\bibinfo {author} {\bibfnamefont {N.}~\bibnamefont
  {Kaloper}}, \bibinfo {author} {\bibfnamefont {A.}~\bibnamefont {Padilla}}, \
  and\ \bibinfo {author} {\bibfnamefont {N.}~\bibnamefont {Tanahashi}},\ }\href
  {\doibase 10.1007/JHEP10(2011)148} {\bibfield  {journal} {\bibinfo  {journal}
  {JHEP}\ }\textbf {\bibinfo {volume} {10}},\ \bibinfo {pages} {148} (\bibinfo
  {year} {2011})},\ \Eprint {http://arxiv.org/abs/1106.4827} {arXiv:1106.4827
  [hep-th]} \BibitemShut {NoStop}%
\bibitem [{\citenamefont {Kimura}\ \emph {et~al.}(2012)\citenamefont {Kimura},
  \citenamefont {Kobayashi},\ and\ \citenamefont {Yamamoto}}]{Kimura:2011dc}%
  \BibitemOpen
  \bibfield  {author} {\bibinfo {author} {\bibfnamefont {R.}~\bibnamefont
  {Kimura}}, \bibinfo {author} {\bibfnamefont {T.}~\bibnamefont {Kobayashi}}, \
  and\ \bibinfo {author} {\bibfnamefont {K.}~\bibnamefont {Yamamoto}},\ }\href
  {\doibase 10.1103/PhysRevD.85.024023} {\bibfield  {journal} {\bibinfo
  {journal} {Phys.Rev.}\ }\textbf {\bibinfo {volume} {D85}},\ \bibinfo {pages}
  {024023} (\bibinfo {year} {2012})},\ \Eprint {http://arxiv.org/abs/1111.6749}
  {arXiv:1111.6749 [astro-ph.CO]} \BibitemShut {NoStop}%
\bibitem [{\citenamefont {Babichev}\ and\ \citenamefont
  {Deffayet}(2013)}]{Babichev:2013usa}%
  \BibitemOpen
  \bibfield  {author} {\bibinfo {author} {\bibfnamefont {E.}~\bibnamefont
  {Babichev}}\ and\ \bibinfo {author} {\bibfnamefont {C.}~\bibnamefont
  {Deffayet}},\ }\href {\doibase 10.1088/0264-9381/30/18/184001} {\bibfield
  {journal} {\bibinfo  {journal} {Class. Quant. Grav.}\ }\textbf {\bibinfo
  {volume} {30}},\ \bibinfo {pages} {184001} (\bibinfo {year} {2013})},\
  \Eprint {http://arxiv.org/abs/1304.7240} {arXiv:1304.7240 [gr-qc]}
  \BibitemShut {NoStop}%
\bibitem [{\citenamefont {Koyama}\ \emph {et~al.}(2013)\citenamefont {Koyama},
  \citenamefont {Niz},\ and\ \citenamefont {Tasinato}}]{Koyama:2013paa}%
  \BibitemOpen
  \bibfield  {author} {\bibinfo {author} {\bibfnamefont {K.}~\bibnamefont
  {Koyama}}, \bibinfo {author} {\bibfnamefont {G.}~\bibnamefont {Niz}}, \ and\
  \bibinfo {author} {\bibfnamefont {G.}~\bibnamefont {Tasinato}},\ }\href
  {\doibase 10.1103/PhysRevD.88.021502} {\bibfield  {journal} {\bibinfo
  {journal} {Phys.Rev.}\ }\textbf {\bibinfo {volume} {D88}},\ \bibinfo {pages}
  {021502} (\bibinfo {year} {2013})},\ \Eprint {http://arxiv.org/abs/1305.0279}
  {arXiv:1305.0279 [hep-th]} \BibitemShut {NoStop}%
\bibitem [{\citenamefont {Horndeski}(1974)}]{Horndeski:1974wa}%
  \BibitemOpen
  \bibfield  {author} {\bibinfo {author} {\bibfnamefont {G.~W.}\ \bibnamefont
  {Horndeski}},\ }\href {\doibase 10.1007/BF01807638} {\bibfield  {journal}
  {\bibinfo  {journal} {Int.J.Theor.Phys.}\ }\textbf {\bibinfo {volume} {10}},\
  \bibinfo {pages} {363} (\bibinfo {year} {1974})}\BibitemShut {NoStop}%
\bibitem [{\citenamefont {Deffayet}\ \emph
  {et~al.}(2009{\natexlab{a}})\citenamefont {Deffayet}, \citenamefont
  {Esposito-Farese},\ and\ \citenamefont {Vikman}}]{Deffayet:2009wt}%
  \BibitemOpen
  \bibfield  {author} {\bibinfo {author} {\bibfnamefont {C.}~\bibnamefont
  {Deffayet}}, \bibinfo {author} {\bibfnamefont {G.}~\bibnamefont
  {Esposito-Farese}}, \ and\ \bibinfo {author} {\bibfnamefont {A.}~\bibnamefont
  {Vikman}},\ }\href {\doibase 10.1103/PhysRevD.79.084003} {\bibfield
  {journal} {\bibinfo  {journal} {Phys.Rev.}\ }\textbf {\bibinfo {volume}
  {D79}},\ \bibinfo {pages} {084003} (\bibinfo {year} {2009}{\natexlab{a}})},\
  \Eprint {http://arxiv.org/abs/0901.1314} {arXiv:0901.1314 [hep-th]}
  \BibitemShut {NoStop}%
\bibitem [{\citenamefont {Deffayet}\ \emph
  {et~al.}(2009{\natexlab{b}})\citenamefont {Deffayet}, \citenamefont {Deser},\
  and\ \citenamefont {Esposito-Farese}}]{Deffayet:2009mn}%
  \BibitemOpen
  \bibfield  {author} {\bibinfo {author} {\bibfnamefont {C.}~\bibnamefont
  {Deffayet}}, \bibinfo {author} {\bibfnamefont {S.}~\bibnamefont {Deser}}, \
  and\ \bibinfo {author} {\bibfnamefont {G.}~\bibnamefont {Esposito-Farese}},\
  }\href {\doibase 10.1103/PhysRevD.80.064015} {\bibfield  {journal} {\bibinfo
  {journal} {Phys. Rev.}\ }\textbf {\bibinfo {volume} {D80}},\ \bibinfo {pages}
  {064015} (\bibinfo {year} {2009}{\natexlab{b}})},\ \Eprint
  {http://arxiv.org/abs/0906.1967} {arXiv:0906.1967 [gr-qc]} \BibitemShut
  {NoStop}%
\bibitem [{\citenamefont {Deffayet}\ \emph {et~al.}(2011)\citenamefont
  {Deffayet}, \citenamefont {Gao}, \citenamefont {Steer},\ and\ \citenamefont
  {Zahariade}}]{Deffayet:2011gz}%
  \BibitemOpen
  \bibfield  {author} {\bibinfo {author} {\bibfnamefont {C.}~\bibnamefont
  {Deffayet}}, \bibinfo {author} {\bibfnamefont {X.}~\bibnamefont {Gao}},
  \bibinfo {author} {\bibfnamefont {D.~A.}\ \bibnamefont {Steer}}, \ and\
  \bibinfo {author} {\bibfnamefont {G.}~\bibnamefont {Zahariade}},\ }\href
  {\doibase 10.1103/PhysRevD.84.064039} {\bibfield  {journal} {\bibinfo
  {journal} {Phys. Rev.}\ }\textbf {\bibinfo {volume} {D84}},\ \bibinfo {pages}
  {064039} (\bibinfo {year} {2011})},\ \Eprint {http://arxiv.org/abs/1103.3260}
  {arXiv:1103.3260 [hep-th]} \BibitemShut {NoStop}%
\bibitem [{\citenamefont {Kaloper}(2004)}]{Kaloper:2003yf}%
  \BibitemOpen
  \bibfield  {author} {\bibinfo {author} {\bibfnamefont {N.}~\bibnamefont
  {Kaloper}},\ }\href {\doibase 10.1016/j.physletb.2004.01.005} {\bibfield
  {journal} {\bibinfo  {journal} {Phys.Lett.}\ }\textbf {\bibinfo {volume}
  {B583}},\ \bibinfo {pages} {1} (\bibinfo {year} {2004})},\ \Eprint
  {http://arxiv.org/abs/hep-ph/0312002} {arXiv:hep-ph/0312002 [hep-ph]}
  \BibitemShut {NoStop}%
\bibitem [{\citenamefont {Kobayashi}\ \emph {et~al.}(2011)\citenamefont
  {Kobayashi}, \citenamefont {Yamaguchi},\ and\ \citenamefont
  {Yokoyama}}]{Kobayashi:2011nu}%
  \BibitemOpen
  \bibfield  {author} {\bibinfo {author} {\bibfnamefont {T.}~\bibnamefont
  {Kobayashi}}, \bibinfo {author} {\bibfnamefont {M.}~\bibnamefont
  {Yamaguchi}}, \ and\ \bibinfo {author} {\bibfnamefont {J.}~\bibnamefont
  {Yokoyama}},\ }\href {\doibase 10.1143/PTP.126.511} {\bibfield  {journal}
  {\bibinfo  {journal} {Prog. Theor. Phys.}\ }\textbf {\bibinfo {volume}
  {126}},\ \bibinfo {pages} {511} (\bibinfo {year} {2011})},\ \Eprint
  {http://arxiv.org/abs/1105.5723} {arXiv:1105.5723 [hep-th]} \BibitemShut
  {NoStop}%
\bibitem [{\citenamefont {De~Felice}\ and\ \citenamefont
  {Tsujikawa}(2012)}]{DeFelice:2011bh}%
  \BibitemOpen
  \bibfield  {author} {\bibinfo {author} {\bibfnamefont {A.}~\bibnamefont
  {De~Felice}}\ and\ \bibinfo {author} {\bibfnamefont {S.}~\bibnamefont
  {Tsujikawa}},\ }\href {\doibase 10.1088/1475-7516/2012/02/007} {\bibfield
  {journal} {\bibinfo  {journal} {JCAP}\ }\textbf {\bibinfo {volume} {1202}},\
  \bibinfo {pages} {007} (\bibinfo {year} {2012})},\ \Eprint
  {http://arxiv.org/abs/1110.3878} {arXiv:1110.3878 [gr-qc]} \BibitemShut
  {NoStop}%
\bibitem [{\citenamefont {De~Felice}\ and\ \citenamefont
  {Tsujikawa}(2010)}]{DeFelice:2010pv}%
  \BibitemOpen
  \bibfield  {author} {\bibinfo {author} {\bibfnamefont {A.}~\bibnamefont
  {De~Felice}}\ and\ \bibinfo {author} {\bibfnamefont {S.}~\bibnamefont
  {Tsujikawa}},\ }\href {\doibase 10.1103/PhysRevLett.105.111301} {\bibfield
  {journal} {\bibinfo  {journal} {Phys.Rev.Lett.}\ }\textbf {\bibinfo {volume}
  {105}},\ \bibinfo {pages} {111301} (\bibinfo {year} {2010})},\ \Eprint
  {http://arxiv.org/abs/1007.2700} {arXiv:1007.2700 [astro-ph.CO]} \BibitemShut
  {NoStop}%
\bibitem [{\citenamefont {De~Felice}\ and\ \citenamefont
  {Tsujikawa}(2011)}]{DeFelice:2010nf}%
  \BibitemOpen
  \bibfield  {author} {\bibinfo {author} {\bibfnamefont {A.}~\bibnamefont
  {De~Felice}}\ and\ \bibinfo {author} {\bibfnamefont {S.}~\bibnamefont
  {Tsujikawa}},\ }\href {\doibase 10.1103/PhysRevD.84.124029} {\bibfield
  {journal} {\bibinfo  {journal} {Phys. Rev.}\ }\textbf {\bibinfo {volume}
  {D84}},\ \bibinfo {pages} {124029} (\bibinfo {year} {2011})},\ \Eprint
  {http://arxiv.org/abs/1008.4236} {arXiv:1008.4236 [hep-th]} \BibitemShut
  {NoStop}%
\bibitem [{\citenamefont {Kase}\ and\ \citenamefont
  {Tsujikawa}(2014)}]{Kase:2014yya}%
  \BibitemOpen
  \bibfield  {author} {\bibinfo {author} {\bibfnamefont {R.}~\bibnamefont
  {Kase}}\ and\ \bibinfo {author} {\bibfnamefont {S.}~\bibnamefont
  {Tsujikawa}},\ }\href {\doibase 10.1103/PhysRevD.90.044073} {\bibfield
  {journal} {\bibinfo  {journal} {Phys.Rev.}\ }\textbf {\bibinfo {volume}
  {D90}},\ \bibinfo {pages} {044073} (\bibinfo {year} {2014})},\ \Eprint
  {http://arxiv.org/abs/1407.0794} {arXiv:1407.0794 [hep-th]} \BibitemShut
  {NoStop}%
\bibitem [{\citenamefont {Bellini}\ and\ \citenamefont
  {Sawicki}(2014)}]{Bellini:2014fua}%
  \BibitemOpen
  \bibfield  {author} {\bibinfo {author} {\bibfnamefont {E.}~\bibnamefont
  {Bellini}}\ and\ \bibinfo {author} {\bibfnamefont {I.}~\bibnamefont
  {Sawicki}},\ }\href {\doibase 10.1088/1475-7516/2014/07/050} {\bibfield
  {journal} {\bibinfo  {journal} {JCAP}\ }\textbf {\bibinfo {volume} {1407}},\
  \bibinfo {pages} {050} (\bibinfo {year} {2014})},\ \Eprint
  {http://arxiv.org/abs/1404.3713} {arXiv:1404.3713 [astro-ph.CO]} \BibitemShut
  {NoStop}%
\bibitem [{\citenamefont {Cisterna}\ \emph {et~al.}(2015)\citenamefont
  {Cisterna}, \citenamefont {Delsate},\ and\ \citenamefont
  {Rinaldi}}]{Cisterna:2015yla}%
  \BibitemOpen
  \bibfield  {author} {\bibinfo {author} {\bibfnamefont {A.}~\bibnamefont
  {Cisterna}}, \bibinfo {author} {\bibfnamefont {T.}~\bibnamefont {Delsate}}, \
  and\ \bibinfo {author} {\bibfnamefont {M.}~\bibnamefont {Rinaldi}},\ }\href
  {\doibase 10.1103/PhysRevD.92.044050} {\bibfield  {journal} {\bibinfo
  {journal} {Phys. Rev.}\ }\textbf {\bibinfo {volume} {D92}},\ \bibinfo {pages}
  {044050} (\bibinfo {year} {2015})},\ \Eprint
  {http://arxiv.org/abs/1504.05189} {arXiv:1504.05189 [gr-qc]} \BibitemShut
  {NoStop}%
\bibitem [{\citenamefont {Maselli}\ \emph {et~al.}(2016)\citenamefont
  {Maselli}, \citenamefont {Silva}, \citenamefont {Minamitsuji},\ and\
  \citenamefont {Berti}}]{Maselli:2016gxk}%
  \BibitemOpen
  \bibfield  {author} {\bibinfo {author} {\bibfnamefont {A.}~\bibnamefont
  {Maselli}}, \bibinfo {author} {\bibfnamefont {H.~O.}\ \bibnamefont {Silva}},
  \bibinfo {author} {\bibfnamefont {M.}~\bibnamefont {Minamitsuji}}, \ and\
  \bibinfo {author} {\bibfnamefont {E.}~\bibnamefont {Berti}},\ }\href
  {\doibase 10.1103/PhysRevD.93.124056} {\bibfield  {journal} {\bibinfo
  {journal} {Phys. Rev.}\ }\textbf {\bibinfo {volume} {D93}},\ \bibinfo {pages}
  {124056} (\bibinfo {year} {2016})},\ \Eprint
  {http://arxiv.org/abs/1603.04876} {arXiv:1603.04876 [gr-qc]} \BibitemShut
  {NoStop}%
\bibitem [{\citenamefont {Babichev}\ \emph
  {et~al.}(2016{\natexlab{a}})\citenamefont {Babichev}, \citenamefont {Koyama},
  \citenamefont {Langlois}, \citenamefont {Saito},\ and\ \citenamefont
  {Sakstein}}]{Babichev:2016jom}%
  \BibitemOpen
  \bibfield  {author} {\bibinfo {author} {\bibfnamefont {E.}~\bibnamefont
  {Babichev}}, \bibinfo {author} {\bibfnamefont {K.}~\bibnamefont {Koyama}},
  \bibinfo {author} {\bibfnamefont {D.}~\bibnamefont {Langlois}}, \bibinfo
  {author} {\bibfnamefont {R.}~\bibnamefont {Saito}}, \ and\ \bibinfo {author}
  {\bibfnamefont {J.}~\bibnamefont {Sakstein}},\ }\href {\doibase
  10.1088/0264-9381/33/23/235014} {\bibfield  {journal} {\bibinfo  {journal}
  {Class. Quant. Grav.}\ }\textbf {\bibinfo {volume} {33}},\ \bibinfo {pages}
  {235014} (\bibinfo {year} {2016}{\natexlab{a}})},\ \Eprint
  {http://arxiv.org/abs/1606.06627} {arXiv:1606.06627 [gr-qc]} \BibitemShut
  {NoStop}%
\bibitem [{\citenamefont {Sakstein}\ \emph
  {et~al.}(2016{\natexlab{a}})\citenamefont {Sakstein}, \citenamefont
  {Babichev}, \citenamefont {Koyama}, \citenamefont {Langlois},\ and\
  \citenamefont {Saito}}]{Sakstein:2016oel}%
  \BibitemOpen
  \bibfield  {author} {\bibinfo {author} {\bibfnamefont {J.}~\bibnamefont
  {Sakstein}}, \bibinfo {author} {\bibfnamefont {E.}~\bibnamefont {Babichev}},
  \bibinfo {author} {\bibfnamefont {K.}~\bibnamefont {Koyama}}, \bibinfo
  {author} {\bibfnamefont {D.}~\bibnamefont {Langlois}}, \ and\ \bibinfo
  {author} {\bibfnamefont {R.}~\bibnamefont {Saito}},\ }\href@noop {} {\
  (\bibinfo {year} {2016}{\natexlab{a}})},\ \Eprint
  {http://arxiv.org/abs/1612.04263} {arXiv:1612.04263 [gr-qc]} \BibitemShut
  {NoStop}%
\bibitem [{\citenamefont {Koyama}\ and\ \citenamefont
  {Sakstein}(2015)}]{Koyama:2015oma}%
  \BibitemOpen
  \bibfield  {author} {\bibinfo {author} {\bibfnamefont {K.}~\bibnamefont
  {Koyama}}\ and\ \bibinfo {author} {\bibfnamefont {J.}~\bibnamefont
  {Sakstein}},\ }\href {\doibase 10.1103/PhysRevD.91.124066} {\bibfield
  {journal} {\bibinfo  {journal} {Phys.Rev.}\ }\textbf {\bibinfo {volume}
  {D91}},\ \bibinfo {pages} {124066} (\bibinfo {year} {2015})},\ \Eprint
  {http://arxiv.org/abs/1502.06872} {arXiv:1502.06872 [astro-ph.CO]}
  \BibitemShut {NoStop}%
\bibitem [{\citenamefont {Saito}\ \emph {et~al.}(2015)\citenamefont {Saito},
  \citenamefont {Yamauchi}, \citenamefont {Mizuno}, \citenamefont {Gleyzes},\
  and\ \citenamefont {Langlois}}]{Saito:2015fza}%
  \BibitemOpen
  \bibfield  {author} {\bibinfo {author} {\bibfnamefont {R.}~\bibnamefont
  {Saito}}, \bibinfo {author} {\bibfnamefont {D.}~\bibnamefont {Yamauchi}},
  \bibinfo {author} {\bibfnamefont {S.}~\bibnamefont {Mizuno}}, \bibinfo
  {author} {\bibfnamefont {J.}~\bibnamefont {Gleyzes}}, \ and\ \bibinfo
  {author} {\bibfnamefont {D.}~\bibnamefont {Langlois}},\ }\href {\doibase
  10.1088/1475-7516/2015/06/008} {\bibfield  {journal} {\bibinfo  {journal}
  {JCAP}\ }\textbf {\bibinfo {volume} {1506}},\ \bibinfo {pages} {008}
  (\bibinfo {year} {2015})},\ \Eprint {http://arxiv.org/abs/1503.01448}
  {arXiv:1503.01448 [gr-qc]} \BibitemShut {NoStop}%
\bibitem [{\citenamefont {Sakstein}\ and\ \citenamefont
  {Koyama}(2015)}]{Sakstein:2015aqx}%
  \BibitemOpen
  \bibfield  {author} {\bibinfo {author} {\bibfnamefont {J.}~\bibnamefont
  {Sakstein}}\ and\ \bibinfo {author} {\bibfnamefont {K.}~\bibnamefont
  {Koyama}},\ }\href {\doibase 10.1142/S0218271815440216} {\bibfield  {journal}
  {\bibinfo  {journal} {Int. J. Mod. Phys.}\ }\textbf {\bibinfo {volume}
  {D24}},\ \bibinfo {pages} {1544021} (\bibinfo {year} {2015})}\BibitemShut
  {NoStop}%
\bibitem [{\citenamefont {Sakstein}(2015{\natexlab{a}})}]{Sakstein:2015zoa}%
  \BibitemOpen
  \bibfield  {author} {\bibinfo {author} {\bibfnamefont {J.}~\bibnamefont
  {Sakstein}},\ }\href {\doibase 10.1103/PhysRevLett.115.201101} {\bibfield
  {journal} {\bibinfo  {journal} {Phys. Rev. Lett.}\ }\textbf {\bibinfo
  {volume} {115}},\ \bibinfo {pages} {201101} (\bibinfo {year}
  {2015}{\natexlab{a}})},\ \Eprint {http://arxiv.org/abs/1510.05964}
  {arXiv:1510.05964 [astro-ph.CO]} \BibitemShut {NoStop}%
\bibitem [{\citenamefont {Sakstein}(2015{\natexlab{b}})}]{Sakstein:2015aac}%
  \BibitemOpen
  \bibfield  {author} {\bibinfo {author} {\bibfnamefont {J.}~\bibnamefont
  {Sakstein}},\ }\href {\doibase 10.1103/PhysRevD.92.124045} {\bibfield
  {journal} {\bibinfo  {journal} {Phys. Rev.}\ }\textbf {\bibinfo {volume}
  {D92}},\ \bibinfo {pages} {124045} (\bibinfo {year} {2015}{\natexlab{b}})},\
  \Eprint {http://arxiv.org/abs/1511.01685} {arXiv:1511.01685 [astro-ph.CO]}
  \BibitemShut {NoStop}%
\bibitem [{\citenamefont {Jain}\ \emph {et~al.}(2015)\citenamefont {Jain},
  \citenamefont {Kouvaris},\ and\ \citenamefont {Nielsen}}]{Jain:2015edg}%
  \BibitemOpen
  \bibfield  {author} {\bibinfo {author} {\bibfnamefont {R.~K.}\ \bibnamefont
  {Jain}}, \bibinfo {author} {\bibfnamefont {C.}~\bibnamefont {Kouvaris}}, \
  and\ \bibinfo {author} {\bibfnamefont {N.~G.}\ \bibnamefont {Nielsen}},\
  }\href@noop {} {\  (\bibinfo {year} {2015})},\ \Eprint
  {http://arxiv.org/abs/1512.05946} {arXiv:1512.05946 [astro-ph.CO]}
  \BibitemShut {NoStop}%
\bibitem [{\citenamefont {Sakstein}\ \emph
  {et~al.}(2016{\natexlab{b}})\citenamefont {Sakstein}, \citenamefont {Wilcox},
  \citenamefont {Bacon}, \citenamefont {Koyama},\ and\ \citenamefont
  {Nichol}}]{Sakstein:2016ggl}%
  \BibitemOpen
  \bibfield  {author} {\bibinfo {author} {\bibfnamefont {J.}~\bibnamefont
  {Sakstein}}, \bibinfo {author} {\bibfnamefont {H.}~\bibnamefont {Wilcox}},
  \bibinfo {author} {\bibfnamefont {D.}~\bibnamefont {Bacon}}, \bibinfo
  {author} {\bibfnamefont {K.}~\bibnamefont {Koyama}}, \ and\ \bibinfo {author}
  {\bibfnamefont {R.~C.}\ \bibnamefont {Nichol}},\ }\href {\doibase
  10.1088/1475-7516/2016/07/019} {\bibfield  {journal} {\bibinfo  {journal}
  {JCAP}\ }\textbf {\bibinfo {volume} {1607}},\ \bibinfo {pages} {019}
  (\bibinfo {year} {2016}{\natexlab{b}})},\ \Eprint
  {http://arxiv.org/abs/1603.06368} {arXiv:1603.06368 [astro-ph.CO]}
  \BibitemShut {NoStop}%
\bibitem [{\citenamefont {Sakstein}\ \emph {et~al.}(2017)\citenamefont
  {Sakstein}, \citenamefont {Kenna-Allison},\ and\ \citenamefont
  {Koyama}}]{Sakstein:2016lyj}%
  \BibitemOpen
  \bibfield  {author} {\bibinfo {author} {\bibfnamefont {J.}~\bibnamefont
  {Sakstein}}, \bibinfo {author} {\bibfnamefont {M.}~\bibnamefont
  {Kenna-Allison}}, \ and\ \bibinfo {author} {\bibfnamefont {K.}~\bibnamefont
  {Koyama}},\ }\href {\doibase 10.1088/1475-7516/2017/03/007} {\bibfield
  {journal} {\bibinfo  {journal} {JCAP}\ }\textbf {\bibinfo {volume} {1703}},\
  \bibinfo {pages} {007} (\bibinfo {year} {2017})},\ \Eprint
  {http://arxiv.org/abs/1611.01062} {arXiv:1611.01062 [gr-qc]} \BibitemShut
  {NoStop}%
\bibitem [{\citenamefont {Bekenstein}(1996)}]{Bekenstein:1996pn}%
  \BibitemOpen
  \bibfield  {author} {\bibinfo {author} {\bibfnamefont {J.~D.}\ \bibnamefont
  {Bekenstein}},\ }in\ \href
  {http://alice.cern.ch/format/showfull?sysnb=0226057} {\emph {\bibinfo
  {booktitle} {{Physics. Proceedings, 2nd International A.D. Sakharov
  Conference, Moscow, Russia, May 20-24, 1996}}}}\ (\bibinfo {year} {1996})\
  pp.\ \bibinfo {pages} {216--219},\ \Eprint
  {http://arxiv.org/abs/gr-qc/9605059} {arXiv:gr-qc/9605059 [gr-qc]}
  \BibitemShut {NoStop}%
\bibitem [{\citenamefont {Sotiriou}\ and\ \citenamefont
  {Faraoni}(2012)}]{Sotiriou:2011dz}%
  \BibitemOpen
  \bibfield  {author} {\bibinfo {author} {\bibfnamefont {T.~P.}\ \bibnamefont
  {Sotiriou}}\ and\ \bibinfo {author} {\bibfnamefont {V.}~\bibnamefont
  {Faraoni}},\ }\href {\doibase 10.1103/PhysRevLett.108.081103} {\bibfield
  {journal} {\bibinfo  {journal} {Phys. Rev. Lett.}\ }\textbf {\bibinfo
  {volume} {108}},\ \bibinfo {pages} {081103} (\bibinfo {year} {2012})},\
  \Eprint {http://arxiv.org/abs/1109.6324} {arXiv:1109.6324 [gr-qc]}
  \BibitemShut {NoStop}%
\bibitem [{\citenamefont {Faraoni}\ and\ \citenamefont
  {Sotiriou}(2015)}]{Faraoni:2013iea}%
  \BibitemOpen
  \bibfield  {author} {\bibinfo {author} {\bibfnamefont {V.}~\bibnamefont
  {Faraoni}}\ and\ \bibinfo {author} {\bibfnamefont {T.~P.}\ \bibnamefont
  {Sotiriou}},\ }in\ \href {\doibase 10.1142/9789814623995_0095} {\emph
  {\bibinfo {booktitle} {{Proceedings, 13th Marcel Grossmann Meeting on Recent
  Developments in Theoretical and Experimental General Relativity,
  Astrophysics, and Relativistic Field Theories (MG13): Stockholm, Sweden, July
  1-7, 2012}}}}\ (\bibinfo {year} {2015})\ pp.\ \bibinfo {pages} {1119--1121},\
  \Eprint {http://arxiv.org/abs/1303.0746} {arXiv:1303.0746 [gr-qc]}
  \BibitemShut {NoStop}%
\bibitem [{\citenamefont {Hui}\ and\ \citenamefont
  {Nicolis}(2013)}]{Hui:2012qt}%
  \BibitemOpen
  \bibfield  {author} {\bibinfo {author} {\bibfnamefont {L.}~\bibnamefont
  {Hui}}\ and\ \bibinfo {author} {\bibfnamefont {A.}~\bibnamefont {Nicolis}},\
  }\href {\doibase 10.1103/PhysRevLett.110.241104} {\bibfield  {journal}
  {\bibinfo  {journal} {Phys.Rev.Lett.}\ }\textbf {\bibinfo {volume} {110}},\
  \bibinfo {pages} {241104} (\bibinfo {year} {2013})},\ \Eprint
  {http://arxiv.org/abs/1202.1296} {arXiv:1202.1296 [hep-th]} \BibitemShut
  {NoStop}%
\bibitem [{\citenamefont {Sotiriou}\ and\ \citenamefont
  {Zhou}(2014{\natexlab{a}})}]{Sotiriou:2013qea}%
  \BibitemOpen
  \bibfield  {author} {\bibinfo {author} {\bibfnamefont {T.~P.}\ \bibnamefont
  {Sotiriou}}\ and\ \bibinfo {author} {\bibfnamefont {S.-Y.}\ \bibnamefont
  {Zhou}},\ }\href {\doibase 10.1103/PhysRevLett.112.251102} {\bibfield
  {journal} {\bibinfo  {journal} {Phys. Rev. Lett.}\ }\textbf {\bibinfo
  {volume} {112}},\ \bibinfo {pages} {251102} (\bibinfo {year}
  {2014}{\natexlab{a}})},\ \Eprint {http://arxiv.org/abs/1312.3622}
  {arXiv:1312.3622 [gr-qc]} \BibitemShut {NoStop}%
\bibitem [{\citenamefont {Sotiriou}\ and\ \citenamefont
  {Zhou}(2014{\natexlab{b}})}]{Sotiriou:2014pfa}%
  \BibitemOpen
  \bibfield  {author} {\bibinfo {author} {\bibfnamefont {T.~P.}\ \bibnamefont
  {Sotiriou}}\ and\ \bibinfo {author} {\bibfnamefont {S.-Y.}\ \bibnamefont
  {Zhou}},\ }\href {\doibase 10.1103/PhysRevD.90.124063} {\bibfield  {journal}
  {\bibinfo  {journal} {Phys. Rev.}\ }\textbf {\bibinfo {volume} {D90}},\
  \bibinfo {pages} {124063} (\bibinfo {year} {2014}{\natexlab{b}})},\ \bibinfo
  {note} {[Phys. Rev.D90,12(2014)]},\ \Eprint {http://arxiv.org/abs/1408.1698}
  {arXiv:1408.1698 [gr-qc]} \BibitemShut {NoStop}%
\bibitem [{\citenamefont {Benkel}\ \emph {et~al.}(2017)\citenamefont {Benkel},
  \citenamefont {Sotiriou},\ and\ \citenamefont {Witek}}]{Benkel:2016rlz}%
  \BibitemOpen
  \bibfield  {author} {\bibinfo {author} {\bibfnamefont {R.}~\bibnamefont
  {Benkel}}, \bibinfo {author} {\bibfnamefont {T.~P.}\ \bibnamefont
  {Sotiriou}}, \ and\ \bibinfo {author} {\bibfnamefont {H.}~\bibnamefont
  {Witek}},\ }\href {\doibase 10.1088/1361-6382/aa5ce7} {\bibfield  {journal}
  {\bibinfo  {journal} {Class. Quant. Grav.}\ }\textbf {\bibinfo {volume}
  {34}},\ \bibinfo {pages} {064001} (\bibinfo {year} {2017})},\ \Eprint
  {http://arxiv.org/abs/1610.09168} {arXiv:1610.09168 [gr-qc]} \BibitemShut
  {NoStop}%
\bibitem [{\citenamefont {Babichev}\ \emph
  {et~al.}(2016{\natexlab{b}})\citenamefont {Babichev}, \citenamefont
  {Charmousis},\ and\ \citenamefont {Lehébel}}]{Babichev:2016rlq}%
  \BibitemOpen
  \bibfield  {author} {\bibinfo {author} {\bibfnamefont {E.}~\bibnamefont
  {Babichev}}, \bibinfo {author} {\bibfnamefont {C.}~\bibnamefont
  {Charmousis}}, \ and\ \bibinfo {author} {\bibfnamefont {A.}~\bibnamefont
  {Lehébel}},\ }\href {\doibase 10.1088/0264-9381/33/15/154002} {\bibfield
  {journal} {\bibinfo  {journal} {Class. Quant. Grav.}\ }\textbf {\bibinfo
  {volume} {33}},\ \bibinfo {pages} {154002} (\bibinfo {year}
  {2016}{\natexlab{b}})},\ \Eprint {http://arxiv.org/abs/1604.06402}
  {arXiv:1604.06402 [gr-qc]} \BibitemShut {NoStop}%
\bibitem [{\citenamefont {Sakstein}(2014)}]{Sakstein:2014isa}%
  \BibitemOpen
  \bibfield  {author} {\bibinfo {author} {\bibfnamefont {J.}~\bibnamefont
  {Sakstein}},\ }\href {\doibase 10.1088/1475-7516/2014/12/012} {\bibfield
  {journal} {\bibinfo  {journal} {JCAP}\ }\textbf {\bibinfo {volume} {1412}},\
  \bibinfo {pages} {012} (\bibinfo {year} {2014})},\ \Eprint
  {http://arxiv.org/abs/1409.1734} {arXiv:1409.1734 [astro-ph.CO]} \BibitemShut
  {NoStop}%
\bibitem [{\citenamefont {Sakstein}(2015{\natexlab{c}})}]{Sakstein:2014aca}%
  \BibitemOpen
  \bibfield  {author} {\bibinfo {author} {\bibfnamefont {J.}~\bibnamefont
  {Sakstein}},\ }\href {\doibase 10.1103/PhysRevD.91.024036} {\bibfield
  {journal} {\bibinfo  {journal} {Phys.Rev.}\ }\textbf {\bibinfo {volume}
  {D91}},\ \bibinfo {pages} {024036} (\bibinfo {year} {2015}{\natexlab{c}})},\
  \Eprint {http://arxiv.org/abs/1409.7296} {arXiv:1409.7296 [astro-ph.CO]}
  \BibitemShut {NoStop}%
\bibitem [{\citenamefont {Ip}\ \emph {et~al.}(2015)\citenamefont {Ip},
  \citenamefont {Sakstein},\ and\ \citenamefont {Schmidt}}]{Ip:2015qsa}%
  \BibitemOpen
  \bibfield  {author} {\bibinfo {author} {\bibfnamefont {H.~Y.}\ \bibnamefont
  {Ip}}, \bibinfo {author} {\bibfnamefont {J.}~\bibnamefont {Sakstein}}, \ and\
  \bibinfo {author} {\bibfnamefont {F.}~\bibnamefont {Schmidt}},\ }\href
  {\doibase 10.1088/1475-7516/2015/10/051} {\bibfield  {journal} {\bibinfo
  {journal} {JCAP}\ }\textbf {\bibinfo {volume} {1510}},\ \bibinfo {pages}
  {051} (\bibinfo {year} {2015})},\ \Eprint {http://arxiv.org/abs/1507.00568}
  {arXiv:1507.00568 [gr-qc]} \BibitemShut {NoStop}%
\bibitem [{\citenamefont {Sakstein}\ and\ \citenamefont
  {Verner}(2015)}]{Sakstein:2015jca}%
  \BibitemOpen
  \bibfield  {author} {\bibinfo {author} {\bibfnamefont {J.}~\bibnamefont
  {Sakstein}}\ and\ \bibinfo {author} {\bibfnamefont {S.}~\bibnamefont
  {Verner}},\ }\href {\doibase 10.1103/PhysRevD.92.123005} {\bibfield
  {journal} {\bibinfo  {journal} {Phys. Rev.}\ }\textbf {\bibinfo {volume}
  {D92}},\ \bibinfo {pages} {123005} (\bibinfo {year} {2015})},\ \Eprint
  {http://arxiv.org/abs/1509.05679} {arXiv:1509.05679 [gr-qc]} \BibitemShut
  {NoStop}%
\bibitem [{\citenamefont {Charmousis}\ \emph {et~al.}(2012)\citenamefont
  {Charmousis}, \citenamefont {Copeland}, \citenamefont {Padilla},\ and\
  \citenamefont {Saffin}}]{Charmousis:2011bf}%
  \BibitemOpen
  \bibfield  {author} {\bibinfo {author} {\bibfnamefont {C.}~\bibnamefont
  {Charmousis}}, \bibinfo {author} {\bibfnamefont {E.~J.}\ \bibnamefont
  {Copeland}}, \bibinfo {author} {\bibfnamefont {A.}~\bibnamefont {Padilla}}, \
  and\ \bibinfo {author} {\bibfnamefont {P.~M.}\ \bibnamefont {Saffin}},\
  }\href {\doibase 10.1103/PhysRevLett.108.051101} {\bibfield  {journal}
  {\bibinfo  {journal} {Phys. Rev. Lett.}\ }\textbf {\bibinfo {volume} {108}},\
  \bibinfo {pages} {051101} (\bibinfo {year} {2012})},\ \Eprint
  {http://arxiv.org/abs/1106.2000} {arXiv:1106.2000 [hep-th]} \BibitemShut
  {NoStop}%
\bibitem [{\citenamefont {Bruneton}\ \emph {et~al.}(2012)\citenamefont
  {Bruneton}, \citenamefont {Rinaldi}, \citenamefont {Kanfon}, \citenamefont
  {Hees}, \citenamefont {Schlogel},\ and\ \citenamefont
  {Fuzfa}}]{Bruneton:2012zk}%
  \BibitemOpen
  \bibfield  {author} {\bibinfo {author} {\bibfnamefont {J.-P.}\ \bibnamefont
  {Bruneton}}, \bibinfo {author} {\bibfnamefont {M.}~\bibnamefont {Rinaldi}},
  \bibinfo {author} {\bibfnamefont {A.}~\bibnamefont {Kanfon}}, \bibinfo
  {author} {\bibfnamefont {A.}~\bibnamefont {Hees}}, \bibinfo {author}
  {\bibfnamefont {S.}~\bibnamefont {Schlogel}}, \ and\ \bibinfo {author}
  {\bibfnamefont {A.}~\bibnamefont {Fuzfa}},\ }\href {\doibase
  10.1155/2012/430694} {\bibfield  {journal} {\bibinfo  {journal} {Adv.
  Astron.}\ }\textbf {\bibinfo {volume} {2012}},\ \bibinfo {pages} {430694}
  (\bibinfo {year} {2012})},\ \Eprint {http://arxiv.org/abs/1203.4446}
  {arXiv:1203.4446 [gr-qc]} \BibitemShut {NoStop}%
\bibitem [{\citenamefont {Rinaldi}(2017)}]{Rinaldi:2016oqp}%
  \BibitemOpen
  \bibfield  {author} {\bibinfo {author} {\bibfnamefont {M.}~\bibnamefont
  {Rinaldi}},\ }\href {\doibase 10.1016/j.dark.2017.02.003} {\bibfield
  {journal} {\bibinfo  {journal} {Phys. Dark Univ.}\ }\textbf {\bibinfo
  {volume} {16}},\ \bibinfo {pages} {14} (\bibinfo {year} {2017})},\ \Eprint
  {http://arxiv.org/abs/1608.03839} {arXiv:1608.03839 [gr-qc]} \BibitemShut
  {NoStop}%
\bibitem [{\citenamefont {Cisterna}\ \emph {et~al.}(2016)\citenamefont
  {Cisterna}, \citenamefont {Delsate}, \citenamefont {Ducobu},\ and\
  \citenamefont {Rinaldi}}]{Cisterna:2016vdx}%
  \BibitemOpen
  \bibfield  {author} {\bibinfo {author} {\bibfnamefont {A.}~\bibnamefont
  {Cisterna}}, \bibinfo {author} {\bibfnamefont {T.}~\bibnamefont {Delsate}},
  \bibinfo {author} {\bibfnamefont {L.}~\bibnamefont {Ducobu}}, \ and\ \bibinfo
  {author} {\bibfnamefont {M.}~\bibnamefont {Rinaldi}},\ }\href {\doibase
  10.1103/PhysRevD.93.084046} {\bibfield  {journal} {\bibinfo  {journal} {Phys.
  Rev.}\ }\textbf {\bibinfo {volume} {D93}},\ \bibinfo {pages} {084046}
  (\bibinfo {year} {2016})},\ \Eprint {http://arxiv.org/abs/1602.06939}
  {arXiv:1602.06939 [gr-qc]} \BibitemShut {NoStop}%
\bibitem [{\citenamefont {Rinaldi}(2012)}]{Rinaldi:2012vy}%
  \BibitemOpen
  \bibfield  {author} {\bibinfo {author} {\bibfnamefont {M.}~\bibnamefont
  {Rinaldi}},\ }\href {\doibase 10.1103/PhysRevD.86.084048} {\bibfield
  {journal} {\bibinfo  {journal} {Phys. Rev.}\ }\textbf {\bibinfo {volume}
  {D86}},\ \bibinfo {pages} {084048} (\bibinfo {year} {2012})},\ \Eprint
  {http://arxiv.org/abs/1208.0103} {arXiv:1208.0103 [gr-qc]} \BibitemShut
  {NoStop}%
\bibitem [{\citenamefont {Babichev}\ and\ \citenamefont
  {Charmousis}(2014)}]{Babichev:2013cya}%
  \BibitemOpen
  \bibfield  {author} {\bibinfo {author} {\bibfnamefont {E.}~\bibnamefont
  {Babichev}}\ and\ \bibinfo {author} {\bibfnamefont {C.}~\bibnamefont
  {Charmousis}},\ }\href {\doibase 10.1007/JHEP08(2014)106} {\bibfield
  {journal} {\bibinfo  {journal} {JHEP}\ }\textbf {\bibinfo {volume} {08}},\
  \bibinfo {pages} {106} (\bibinfo {year} {2014})},\ \Eprint
  {http://arxiv.org/abs/1312.3204} {arXiv:1312.3204 [gr-qc]} \BibitemShut
  {NoStop}%
\bibitem [{\citenamefont {Anabalon}\ \emph {et~al.}(2014)\citenamefont
  {Anabalon}, \citenamefont {Cisterna},\ and\ \citenamefont
  {Oliva}}]{Anabalon:2013oea}%
  \BibitemOpen
  \bibfield  {author} {\bibinfo {author} {\bibfnamefont {A.}~\bibnamefont
  {Anabalon}}, \bibinfo {author} {\bibfnamefont {A.}~\bibnamefont {Cisterna}},
  \ and\ \bibinfo {author} {\bibfnamefont {J.}~\bibnamefont {Oliva}},\ }\href
  {\doibase 10.1103/PhysRevD.89.084050} {\bibfield  {journal} {\bibinfo
  {journal} {Phys. Rev.}\ }\textbf {\bibinfo {volume} {D89}},\ \bibinfo {pages}
  {084050} (\bibinfo {year} {2014})},\ \Eprint {http://arxiv.org/abs/1312.3597}
  {arXiv:1312.3597 [gr-qc]} \BibitemShut {NoStop}%
\bibitem [{\citenamefont
  {Minamitsuji}(2014{\natexlab{a}})}]{Minamitsuji:2013ura}%
  \BibitemOpen
  \bibfield  {author} {\bibinfo {author} {\bibfnamefont {M.}~\bibnamefont
  {Minamitsuji}},\ }\href {\doibase 10.1103/PhysRevD.89.064017} {\bibfield
  {journal} {\bibinfo  {journal} {Phys. Rev.}\ }\textbf {\bibinfo {volume}
  {D89}},\ \bibinfo {pages} {064017} (\bibinfo {year} {2014}{\natexlab{a}})},\
  \Eprint {http://arxiv.org/abs/1312.3759} {arXiv:1312.3759 [gr-qc]}
  \BibitemShut {NoStop}%
\bibitem [{\citenamefont {Maldacena}(1999)}]{Maldacena:1997re}%
  \BibitemOpen
  \bibfield  {author} {\bibinfo {author} {\bibfnamefont {J.~M.}\ \bibnamefont
  {Maldacena}},\ }\href {\doibase 10.1023/A:1026654312961} {\bibfield
  {journal} {\bibinfo  {journal} {Int. J. Theor. Phys.}\ }\textbf {\bibinfo
  {volume} {38}},\ \bibinfo {pages} {1113} (\bibinfo {year} {1999})},\ \bibinfo
  {note} {[Adv. Theor. Math. Phys.2,231(1998)]},\ \Eprint
  {http://arxiv.org/abs/hep-th/9711200} {arXiv:hep-th/9711200 [hep-th]}
  \BibitemShut {NoStop}%
\bibitem [{\citenamefont {Gubser}\ \emph {et~al.}(1998)\citenamefont {Gubser},
  \citenamefont {Klebanov},\ and\ \citenamefont {Polyakov}}]{Gubser:1998bc}%
  \BibitemOpen
  \bibfield  {author} {\bibinfo {author} {\bibfnamefont {S.~S.}\ \bibnamefont
  {Gubser}}, \bibinfo {author} {\bibfnamefont {I.~R.}\ \bibnamefont
  {Klebanov}}, \ and\ \bibinfo {author} {\bibfnamefont {A.~M.}\ \bibnamefont
  {Polyakov}},\ }\href {\doibase 10.1016/S0370-2693(98)00377-3} {\bibfield
  {journal} {\bibinfo  {journal} {Phys. Lett.}\ }\textbf {\bibinfo {volume}
  {B428}},\ \bibinfo {pages} {105} (\bibinfo {year} {1998})},\ \Eprint
  {http://arxiv.org/abs/hep-th/9802109} {arXiv:hep-th/9802109 [hep-th]}
  \BibitemShut {NoStop}%
\bibitem [{\citenamefont {Witten}(1998)}]{Witten:1998qj}%
  \BibitemOpen
  \bibfield  {author} {\bibinfo {author} {\bibfnamefont {E.}~\bibnamefont
  {Witten}},\ }\href@noop {} {\bibfield  {journal} {\bibinfo  {journal} {Adv.
  Theor. Math. Phys.}\ }\textbf {\bibinfo {volume} {2}},\ \bibinfo {pages}
  {253} (\bibinfo {year} {1998})},\ \Eprint
  {http://arxiv.org/abs/hep-th/9802150} {arXiv:hep-th/9802150 [hep-th]}
  \BibitemShut {NoStop}%
\bibitem [{\citenamefont {Chan}\ and\ \citenamefont
  {Mann}(1997)}]{Chan:1996yk}%
  \BibitemOpen
  \bibfield  {author} {\bibinfo {author} {\bibfnamefont {J.~S.~F.}\
  \bibnamefont {Chan}}\ and\ \bibinfo {author} {\bibfnamefont {R.~B.}\
  \bibnamefont {Mann}},\ }\href {\doibase 10.1103/PhysRevD.55.7546} {\bibfield
  {journal} {\bibinfo  {journal} {Phys. Rev.}\ }\textbf {\bibinfo {volume}
  {D55}},\ \bibinfo {pages} {7546} (\bibinfo {year} {1997})},\ \Eprint
  {http://arxiv.org/abs/gr-qc/9612026} {arXiv:gr-qc/9612026 [gr-qc]}
  \BibitemShut {NoStop}%
\bibitem [{\citenamefont {Chan}\ and\ \citenamefont
  {Mann}(1999)}]{Chan:1999sc}%
  \BibitemOpen
  \bibfield  {author} {\bibinfo {author} {\bibfnamefont {J.~S.~F.}\
  \bibnamefont {Chan}}\ and\ \bibinfo {author} {\bibfnamefont {R.~B.}\
  \bibnamefont {Mann}},\ }\href {\doibase 10.1103/PhysRevD.59.064025}
  {\bibfield  {journal} {\bibinfo  {journal} {Phys. Rev.}\ }\textbf {\bibinfo
  {volume} {D59}},\ \bibinfo {pages} {064025} (\bibinfo {year}
  {1999})}\BibitemShut {NoStop}%
\bibitem [{\citenamefont {Aharony}\ \emph {et~al.}(2000)\citenamefont
  {Aharony}, \citenamefont {Gubser}, \citenamefont {Maldacena}, \citenamefont
  {Ooguri},\ and\ \citenamefont {Oz}}]{Aharony:1999ti}%
  \BibitemOpen
  \bibfield  {author} {\bibinfo {author} {\bibfnamefont {O.}~\bibnamefont
  {Aharony}}, \bibinfo {author} {\bibfnamefont {S.~S.}\ \bibnamefont {Gubser}},
  \bibinfo {author} {\bibfnamefont {J.~M.}\ \bibnamefont {Maldacena}}, \bibinfo
  {author} {\bibfnamefont {H.}~\bibnamefont {Ooguri}}, \ and\ \bibinfo {author}
  {\bibfnamefont {Y.}~\bibnamefont {Oz}},\ }\href {\doibase
  10.1016/S0370-1573(99)00083-6} {\bibfield  {journal} {\bibinfo  {journal}
  {Phys. Rept.}\ }\textbf {\bibinfo {volume} {323}},\ \bibinfo {pages} {183}
  (\bibinfo {year} {2000})},\ \Eprint {http://arxiv.org/abs/hep-th/9905111}
  {arXiv:hep-th/9905111 [hep-th]} \BibitemShut {NoStop}%
\bibitem [{\citenamefont {Horowitz}\ and\ \citenamefont
  {Hubeny}(2000)}]{Horowitz:1999jd}%
  \BibitemOpen
  \bibfield  {author} {\bibinfo {author} {\bibfnamefont {G.~T.}\ \bibnamefont
  {Horowitz}}\ and\ \bibinfo {author} {\bibfnamefont {V.~E.}\ \bibnamefont
  {Hubeny}},\ }\href {\doibase 10.1103/PhysRevD.62.024027} {\bibfield
  {journal} {\bibinfo  {journal} {Phys. Rev.}\ }\textbf {\bibinfo {volume}
  {D62}},\ \bibinfo {pages} {024027} (\bibinfo {year} {2000})},\ \Eprint
  {http://arxiv.org/abs/hep-th/9909056} {arXiv:hep-th/9909056 [hep-th]}
  \BibitemShut {NoStop}%
\bibitem [{\citenamefont {Konoplya}\ and\ \citenamefont
  {Zhidenko}(2011)}]{Konoplya:2011qq}%
  \BibitemOpen
  \bibfield  {author} {\bibinfo {author} {\bibfnamefont {R.~A.}\ \bibnamefont
  {Konoplya}}\ and\ \bibinfo {author} {\bibfnamefont {A.}~\bibnamefont
  {Zhidenko}},\ }\href {\doibase 10.1103/RevModPhys.83.793} {\bibfield
  {journal} {\bibinfo  {journal} {Rev. Mod. Phys.}\ }\textbf {\bibinfo {volume}
  {83}},\ \bibinfo {pages} {793} (\bibinfo {year} {2011})},\ \Eprint
  {http://arxiv.org/abs/1102.4014} {arXiv:1102.4014 [gr-qc]} \BibitemShut
  {NoStop}%
\bibitem [{\citenamefont {Kord~Zangeneh}\ \emph {et~al.}(2017)\citenamefont
  {Kord~Zangeneh}, \citenamefont {Wang}, \citenamefont {Sheykhi},\ and\
  \citenamefont {Tang}}]{Zangeneh:2017rhc}%
  \BibitemOpen
  \bibfield  {author} {\bibinfo {author} {\bibfnamefont {M.}~\bibnamefont
  {Kord~Zangeneh}}, \bibinfo {author} {\bibfnamefont {B.}~\bibnamefont {Wang}},
  \bibinfo {author} {\bibfnamefont {A.}~\bibnamefont {Sheykhi}}, \ and\
  \bibinfo {author} {\bibfnamefont {Z.~Y.}\ \bibnamefont {Tang}},\ }\href
  {\doibase 10.1016/j.physletb.2017.05.050} {\bibfield  {journal} {\bibinfo
  {journal} {Phys. Lett.}\ }\textbf {\bibinfo {volume} {B771}},\ \bibinfo
  {pages} {257} (\bibinfo {year} {2017})},\ \Eprint
  {http://arxiv.org/abs/1701.03644} {arXiv:1701.03644 [hep-th]} \BibitemShut
  {NoStop}%
\bibitem [{\citenamefont
  {Minamitsuji}(2014{\natexlab{b}})}]{Minamitsuji:2014hha}%
  \BibitemOpen
  \bibfield  {author} {\bibinfo {author} {\bibfnamefont {M.}~\bibnamefont
  {Minamitsuji}},\ }\href {\doibase 10.1007/s10714-014-1785-0} {\bibfield
  {journal} {\bibinfo  {journal} {Gen. Rel. Grav.}\ }\textbf {\bibinfo {volume}
  {46}},\ \bibinfo {pages} {1785} (\bibinfo {year} {2014}{\natexlab{b}})},\
  \Eprint {http://arxiv.org/abs/1407.4901} {arXiv:1407.4901 [gr-qc]}
  \BibitemShut {NoStop}%
\bibitem [{\citenamefont {Konoplya}(2002)}]{Konoplya:2002zu}%
  \BibitemOpen
  \bibfield  {author} {\bibinfo {author} {\bibfnamefont {R.~A.}\ \bibnamefont
  {Konoplya}},\ }\href {\doibase 10.1103/PhysRevD.66.044009} {\bibfield
  {journal} {\bibinfo  {journal} {Phys. Rev.}\ }\textbf {\bibinfo {volume}
  {D66}},\ \bibinfo {pages} {044009} (\bibinfo {year} {2002})},\ \Eprint
  {http://arxiv.org/abs/hep-th/0205142} {arXiv:hep-th/0205142 [hep-th]}
  \BibitemShut {NoStop}%
\end{thebibliography}%

\appendix

\section{Numerical Procedure}
\label{sec:numerical}
In what follows, we review the numerical procedure of Horrowitz and Hubeny \cite{Horowitz:1999jd}. Defining $\Psi=\psi e^{i\omega r^*}$, we can write Eq. (\ref{schrodinger}) in terms of $\Psi$ using the $r$ coordinate so that
\be
f\frac{d^2\Psi}{dr^2}+[f'-2i\omega]\frac{d\Psi}{dr}-U(r)\Psi=0,\label{psi-r}
\ee
where $U(r)=V(r)/f(r)$.  Introducing the new variable $x=1/r$, Eq. (\ref{psi-r}) can be written as 
\be
s(x)\frac{d^2\Psi}{dx^2}+\frac{t(x)}{x-x_h}\frac{d\Psi}{dx}+\frac{u(x)}{(x-x_h)^2}\Psi=0.\label{psi-x}
\ee
Here $x_h=1/r_h$, $s(x)=\frac{x^4f(r)}{x-x_h}$, $t(x)=2x^3f(r)-x^2f'(r)+2i\omega x^2$ and $u(x)=-(x-x_h)U(r)$. From our boundary conditions (Eq. (\ref{bc})), $\Psi$ should be finite as $x\rightarrow x_h$, and vanish as $x\rightarrow 0$. Expanding $\Psi$ as
\be
\Psi=\sum^\infty_{n=0}a_n(x-x_h)^{n+\alpha},\label{psi-series}
\ee
where $\alpha$ is a constant.  We can solve perturbatively by matching every order of the series in $(x-x_h)$. The lowest-order term is 
\be
s_0\alpha(\alpha-1)+t_0\alpha=0,
\ee
where $s_0=-x_h^2f'(r_h)$ and $t_0=-x_h^2f'(r_h)+2i\omega x_h^2$ are the zeroth-order term in the expansion of $s(x)$ and $t(x)$ respectively. There are two solutions given by
\be
\alpha=0, \frac{2i\omega}{f'(r_h)};
\ee
the former corresponds to an incoming wave at the BH horizon, and the latter an outgoing wave there. Physically, as measured by an observer at the horizon, a wave can only travel into the BH, and not out. Therefore, we choose the former solution, i.e. $\alpha=0$. Next, we need to satisfy the other boundary condition, i.e. $\Psi(x=0)=0$. This is achieved by solving the series in $a_n$ to order N, and setting
\be\label{psi0-series}
\Psi(x=0)=\sum^N_{n=0}a_n(\omega)(0-x_h)^n=0.
\ee
We solve the resulting polynomial equation in $\omega$ numerically; the precision of the solution can be checked by varying N and checking the convergence of the results.

\section{Data tables}
In this appendix, we tabulate the QNMs for typical values of the BH horizon radii $r_h$, for $z=1/3$ (SAdS BH). Table \ref{tab-w012} shows the first three QNMs for radial perturbations ($j=0$). Table \ref{tab-wl30} shows the principal QNMs for $j$ up to 30.

\begin{table}[ht]
\centering
{\footnotesize
\def\tabularxcolumn#1{m{#1}}
\begin{tabularx}{\linewidth}{@{}cXX@{}}
\begin{tabular}{c*{3}{c}}
$r_h$   & $\omega_{ST}^{(0)}$ & $\omega_{ST}^{(1)}$ & $\omega_{ST}^{(2)}$ \\
\hline
10 & $ 18.607-26.642i  $ & $ 31.802-49.182i  $ & $ 44.910-71.706i$ \\
8  & $ 14.936-21.315i  $ & $ 25.526-39.353i  $ & $ 36.048-57.379i$ \\
6  & $ 11.284-15.988i  $ & $ 19.282-29.527i  $ & $ 27.229-43.056i$ \\
4  & $ 7.676-10.663i   $ & $ 13.112-19.708i  $ & $ 18.516-28.746i$ \\
2  & $ 4.234-5.340i    $ & $ 7.222-9.911i   $ & $ 10.197-14.476i $ \\
1  & $ 2.798-2.671i    $ & $ 4.758-5.038i    $ & $ 6.719-7.395i  $ \\
0.8 & $2.588-2.130i    $ & $ 4.395-4.062i    $ & $ 6.207-5.983i  $ \\
0.6 & $2.432-1.580i    $ & $ 4.120-3.080i    $ & $ 5.819-4.565i  $ \\
0.4 & $2.363-1.006i    $ & $ 3.979-2.073i    $ & $ 5.617-3.120i  $ \\
0.3 & $2.384-0.704i    $ & $ 3.984-1.547i    $ & $ 5.619-2.369i  $ \\
0.25 & $2.418-0.548i   $ & $ 4.014-1.273i    $ & $ 5.657-1.980i  $ \\
\end{tabular}
\end{tabularx}
\caption{The principal QNM ($\omega_{ST}^{(0)}$) and the first two overtones ($\omega_{ST}^{(1,2)}$), for different BH horizon radii. Here $z=1/3$ and $j=0$. 
\label{tab-w012}
}}
\end{table}

\begin{table}[ht]
\centering
{\footnotesize
\def\tabularxcolumn#1{m{#1}}
\begin{tabularx}{\linewidth}{@{}cXX@{}}
\begin{tabular}{c*{4}{c}}
$r_h$   & $j=0$ & $j=10$ & $j=20$ & $j=30$\\
\hline\\
10 & $ 18.607-26.642i  $ & $ 22.513-25.616i  $ & $ 30.219-24.030i $ & $ 39.098-22.714i $ \\\\
8  & $ 14.936-21.315i $ & $ 19.505-20.155i $ & $ 27.786-18.647i $ & $ 36.979-17.519i $ \\\\
6  & $ 11.284-15.988i $ & $ 16.722-14.691i $ & $ 25.569-13.367i $ & $ 35.043-12.477i $ \\\\
4  & $ 7.676-10.663i $ & $ 14.261-9.273i  $ & $ 23.621-8.265i $ & $ 33.332-7.650i $ \\
\end{tabular}
\end{tabularx}
\caption{The principal QNM for the $z=1/3$ case (SAdS BH), for different $j$-degrees and BH horizon radii $r_h$. \label{tab-wl30}
}}
\end{table}

\end{document}